\documentstyle[12pt,epsf,psfig,eqsection,cite]{article}

\def\thefootnote{\fnsymbol{footnote}}

\newcommand{\eq}{\begin{equation}}
\newcommand{\en}{\end{equation}}
\newcommand{\eqa}{\begin{eqnarray}}
\newcommand{\ena}{\end{eqnarray}}

\newcommand{\br}{\langle}
\newcommand{\kt}{\rangle}

\def\spose#1{\hbox to 0pt{#1\hss}}
\def\ltapprox{\mathrel{\spose{\lower 3pt\hbox{$\mathchar"218$}}
 \raise 2.0pt\hbox{$\mathchar"13C$}}}
\def\gtapprox{\mathrel{\spose{\lower 3pt\hbox{$\mathchar"218$}}
 \raise 2.0pt\hbox{$\mathchar"13E$}}}

\def\case#1#2{{\textstyle\frac{#1}{#2}}}

\newcommand{\PR}[1]{Phys.\ Rev.\ {\bf #1}}

\begin{document}
\begin{titlepage}
\vskip0.5cm
\begin{flushright}
DFTT 39/2000\\
HU-EP-00/29\\
IFUP-TH 999/2000\\
Roma1-1298/00\\
\end{flushright}
\vskip0.5cm
\begin{center}
{\Large\bf The critical equation of state}
\vskip 0.3cm
{\Large\bf of the 2D Ising model}
\end{center}
\vskip 1.3cm
\centerline{
Michele Caselle$^a$, Martin Hasenbusch$^b$, Andrea Pelissetto$^c$, and
Ettore Vicari$^d$}

 \vskip 0.4cm
 \centerline{\sl  $^a$ Dipartimento di Fisica Teorica dell'Universit\`a di
 Torino and I.N.F.N., I-10125 Torino, Italy}
 \centerline{\sl $^b$ Humboldt-Universit\"at zu Berlin, Institut f\"ur Physik,
 Invalidenstr. 110, D-10115 Berlin, Germany}
 \centerline{\sl  $^c$ Dipartimento di Fisica dell'Universit\`a di Roma I
 and I.N.F.N., I-00185 Roma, Italy}
 \centerline{\sl  $^d$ Dipartimento di Fisica dell'Universit\`a di Pisa
 and I.N.F.N., I-56127 Pisa, Italy}
 \medskip
 \centerline{\tt E-mail: caselle@to.infn.it,
                         hasenbus@physik.hu-berlin.de,}
 \centerline{\tt         Andrea.Pelissetto@roma1.infn.it,
                         Vicari@df.unipi.it}
 \vskip 1.cm

\begin{abstract}
We compute the $2n$-point coupling constants in the high-temperature phase of
 the 2d Ising model by using transfer-matrix techniques. This provides
 the first few terms of the expansion of  the effective potential (Helmholtz
 free energy) and of the equation of state
 in terms of the renormalized magnetization.
 By means of a suitable parametric representation,
 we determine an analytic extension of these expansions,
providing the equation of state in the whole critical region
in the $t,h$ plane.

\end{abstract}
\end{titlepage}

\setcounter{footnote}{0}
\def\thefootnote{\arabic{footnote}}

\section{Introduction}
\label{seceintro}

Despite its apparent simplicity and the fact that since more than fifty years
the exact expression of the free energy along the $h=0$ axis is exactly known,
the two-dimensional Ising model still provides many interesting issues.
Since the original Onsager solution~\cite{onsager},
several other exact results have been
obtained for this model. In particular, closed expressions for the two-point
correlation function at $h=0$
and an S-matrix solution on the $t=0,h\not=0$ axis~\cite{zam} exist
(for a review, see~\cite{mccoy}).
However, little is known for generic values of $t$ and $h$ and in
particular, there is no exact expression
for the free energy and for the critical equation of state
in the whole $(t,h)$ plane.

In order to determine the Helmholtz free energy
(also called effective potential)
and the equation of state, we begin by computing the first terms of their
expansion in powers of the magnetization in the high-temperature (HT) phase.
The coefficients of this expansion are directly related to
the $n$-point zero-momentum renormalized couplings $g_n$ that
are also interesting in themselves, since they summarize
relevant (zero-momentum) physical properties of the quantum field
theory that describes the Ising model in the vicinity of the
critical point.
For this purpose we combine transfer matrix (TM)
techniques, conformal field theory (CFT) methods, and general
renormalization-group (RG) properties of critical systems.
Recent works~\cite{ch99,CHPV-00} have shown that
a  combination of these approaches can lead to very accurate
estimates of universal ratios in the 2d Ising model.
In this respect this paper is the natural continuation of ~\cite{CHPV-00} in
which the first nontrivial coupling $g_4$ was determined with the same
techniques used here.

Starting from the expansion of the free energy in powers of the magnetization
in the HT phase, we
determine  approximate representations of the equation of state that
are valid in the critical regime in the whole $(t,h)$ plane. This requires
an analytic continuation in the
complex $t$-plane~\cite{ZJbook,GZ-97}, extending the expansion valid for
$t>0$ to the low-temperature phase $t<0$. For this purpose, we  use  parametric
representations \cite{Schofield-69,SLH-69,Josephson-69},
which implement in a rather simple way the known
analytic properties of the equation of state (Griffiths' analyticity).
We construct a systematic approximation
scheme based on polynomial parametric representations~\cite{GZ-97}
and on a global stationarity condition~\cite{CPRV-99}.
This approach was successfully applied to the three-dimensional Ising model
leading to an accurate determination of
the critical equation of state and of the universal amplitude ratios
that can be obtained from it~\cite{GZ-97,CPRV-99}.

The paper is organized as follows:
In Sec. 2 we set up the formalism and define the quantities that
parametrize the free energy for small values of the magnetization.
These coefficients are evaluated numerically in Sec. 3 by using TM techniques.
In Sec. 4 we use a parametric representation to analytically extend
this expansion to the whole critical region, obtaining
approximate, but still quite precise, expressions of
the critical equation of state in the whole $(t,h)$ plane.
Finally in Sec. 5 we draw our conclusions.
In App. A we report our notations for the universal amplitude
ratios, in App. B we present a detailed discussion of the finite-size behaviour
of the free energy and of its derivatives, and in App. C
we explicitly compute $e_h$, one of the coefficients appearing
in the expansion of the nonlinear scaling field associated with
the magnetic field.

\section{Small-field expansion of the effective potential in the
high-temperature phase}
\label{sece1}

In the theory of critical phenomena continuous phase transitions can be
classified into universality
classes determined only by a few basic properties
characterizing the system, such
as space dimensionality,  range of interaction,
number of components and symmetry of the order parameter.
RG theory predicts that, within a given
universality class, the critical exponents and the scaling functions are
the same for all systems. Here we consider the two-dimensional Ising
universality class, which is characterized by a real order parameter and
effective short-range interactions.
A representative of this universality class is the standard square-lattice
Ising model defined by the partition function
\eq
Z=\sum_{\sigma_i=\pm1}e^{\beta\sum_{\br n,m \kt}\sigma_n\sigma_m
+h\sum_n\sigma_n} ,
\label{zz1}
\en
where the field variable $\sigma_n$ takes the values $\{\pm 1\}$,
$n\equiv(n_0,n_1)$ labels the sites of a square lattice of size $L_0\times L_1$,
and $\br n,m \kt$
denotes a lattice link connecting two nearest-neighbour sites.
In our calculations
we treat asymmetrically the two directions. We denote
by $n_0$ the ``time'' coordinate and by $n_1$ the ``space" one.
We indicate by $N\equiv L_0 L_1$ the number of sites  of
the lattice and define the reduced temperature 
\eq
t\equiv \frac{\beta_c-\beta}{\beta_c} ,
\en
where $\beta_c = \log{(\sqrt{2}+1)}/2 \approx 0.4406868\ldots$ is
the critical point.

We introduce the Gibbs free-energy density 
\eq
F(t,h)\equiv {1\over N} \log(Z(t,h)),
\en
and the related Helmholtz free-energy density---in the field-theoretical
framework usually called effective potential---
\begin{equation}
{\cal F} (t,M) = M\,h - F(t,h),
\end{equation}
where $M$ is the magnetization per site.
For $t>0$, $F(t,h)$ and ${\cal F} (t,M)$ have regular expansions
in even powers of $h$ and $M$ respectively. Explicitly
\begin{eqnarray}
F(t,h) - F(t,0) &=& \sum_{n=1} {1\over (2n)!} \chi_{2n}(t) h^{2n}, \\
{\cal F} (t,M) - {\cal F} (t,0) &=&
      \sum_{n=1} {1\over (2n)!} \chi_{2n}^{\rm 1PI} (t) M^{2n},
\label{expansionFtM}
\end{eqnarray}
where $\chi_{2n}(t)$ is the zero-momentum $n$-point function and
$\chi_{2n}^{\rm 1PI}(t)$ is its one-particle irreducible counterpart. Note that
$\chi_2(t) = \chi(t)$, where $\chi(t)$ is the magnetic susceptibility.

Expansion (\ref{expansionFtM}) can also be written in the equivalent forms
\begin{eqnarray}
{\cal F} (t,M) - {\cal F} (t,0) &=&
\xi_{\rm 2nd}^{-2} \left( {1\over 2} \varphi^2 +
\sum_{n=2} {1\over (2n)!} g_{2n} \varphi^{2n} \right)
\label{freeeng} \\
&=&
 - {\chi^2\over \chi_4} \left(
  {1\over 2} z^2 + {1\over 4!} z^4 + \sum_{n=3} {1\over (2n)!}
   r_{2n} z^{2n} \right),
\label{AZ}
\end{eqnarray}
where
\begin{eqnarray}
\varphi^2 = {\xi_{\rm 2nd}(t)^2 \over \chi(t)} M^2, \qquad \qquad
z^2 = - {\chi_4(t)\over \chi(t)^3} M^2,
\end{eqnarray}
and $\xi_{\rm 2nd}$ is the second-moment correlation length
\eq
\xi_{\rm 2nd} = {1\over 4\chi}
\sum_x x^2 \langle s(0) s(x) \rangle ~~.
\en
The coefficients $g_{2n}$ correspond to the $2n$-point renormalized
coupling constants and $r_{2n} = g_{2n}/g_4^{n-1}$. The couplings
$g_{2n}$ and the related $r_{2n}$ can be computed in
terms of $\chi_{2n}$ and $\xi_{\rm 2nd}$. The four-point coupling is
given by
\eq
g_4(t)=-\frac{\chi_4}{\chi^2\xi_{\rm 2nd}^2}.
\en
An explicit expression of $r_{2n}$ in terms of
$\chi_{2n}$ is given at the beginning of Sec. \ref{sece2}.

The advantage of the expansions (\ref{freeeng}) and (\ref{AZ}) is the
fact that the coefficients have a finite limit---in the following
we indicate it by the same symbol--- for $t\to 0$. Notice
that for $t\to 0$, $z \sim M t^{-\beta}$ and $\chi^2/\chi_4\sim t^{2\nu}$,
so that Eq. (\ref{AZ}) defines the HT expansion of the scaling part
of the free energy
${\cal F}(t,M) \sim t^{2\nu} {\cal F}_{\rm scal}(M t^{-\beta})$.

\section{Calculation of $\lowercase{r}_{2\lowercase{n}}$ with
 transfer-matrix techniques}
\label{sece2}

In this section we wish to report the calculation of the first few
coefficients $r_{2n}$ at the critical point $t=0$.
It is easy to express them in terms
of the critical amplitudes of the $2n$-point functions $\chi_{2n}(t)$.
Defining
\eq
C^+_n =\lim_{t\to0} \chi_n(t)\, t^{15 n/8 - 2},
\label{defCpiun}
\en
it is possible to show by direct calculation that
\begin{eqnarray}
r_6 &=& 10-\frac{C^+_{6} C^+_{2}}{(C^+_{4})^2},
\label{defr6}
\\
r_8 &=& 280-56\frac{C^+_{6} C^+_{2}}{(C^+_{4})^2}
+\frac{C^+_{8} (C^+_{2})^2}{(C^+_{4})^3},
\label{defr8}
\\
r_{10}&=& 15400-4620\frac{C^+_{6} C^+_{2}}{(C^+_{4})^2}
+120 \frac{C^+_{8} (C^+_{2})^2}{(C^+_{4})^3}
\nonumber \\
&&+126 \frac{(C^+_{6})^2 (C^+_{2})^2}{(C^+_{4})^4}
- \frac{C^+_{10} (C^+_{2})^3}{(C^+_{4})^4},
\label{defr10}
\\
r_{12}&=& 1401400-560560\frac{C^+_{6} C^+_{2}}{(C^+_{4})^2}
+17160 \frac{C^+_{8} (C^+_{2})^2}{(C^+_{4})^3}
+36036 \frac{(C^+_{6})^2 (C^+_{2})^3}{(C^+_{4})^4}
\nonumber \\
&& -220 \frac{C^+_{10} (C^+_{2})^3}{(C^+_{4})^4}
   -792 \frac{C^+_{8} C^+_{6} (C^+_{2})^3}{(C^+_{4})^5}
   + \frac{C^+_{12} (C^+_{2})^4}{(C^+_{4})^5}.
\label{defr12}
\ena
The aim of this section is to estimate the constants $C^+_{2n}$, by using
 TM techniques and the exact knowledge of several terms of the small-$t$
 expansion of $\chi_{2n}(t)$. Using the equations reported above, we will
 then obtain our estimates of $r_{2n}$.

\subsection{The transfer-matrix approach}

We obtained our
 numerical estimates of the $2n$-point functions
$\chi_{2n}(t)$ following a two-step procedure. First, by using TM
techniques, we obtained
estimates of these quantities on lattices of finite transverse
size $L_1$, with $L_1\le 24$. Second, we extrapolated these results to the
thermodynamic limit by using the exact knowledge of their Finite Size Scaling
(FSS in the following) behaviour. To perform this second part of the analysis
we have elaborated a new extrapolation scheme which is rather interesting in
itself. We shall discuss it in detail in the last part of this section.

Let us now see both these steps in more detail.
\subsubsection{The TM computation}
This part of the procedure was already discussed  in
Ref.~\cite{CHPV-00}. We report here the main features of the algorithm for
completeness and refer to~\cite{CHPV-00} for a more detailed discussion.

The main idea is to use TM techniques to extract
 the $h$-dependence of the magnetization
at fixed $t$.
To this end, we computed numerically
the magnetization $M(t,h,L_1)$
on lattices of size $\infty \; \times \; L_1$, for $L_1\le 24$.
In this work we use again the data obtained in
Ref.~\cite{CHPV-00}. In addition, we generated
new data for $\beta=0.335$ and $0.345$ and added $L_1=24$
results for $\beta=0.35$ and $0.355$. All computations were performed with
double-precision floating-point arithmetic on commercial workstations.
Typically we obtained $M$ with 15 correct digits.

Then, for each given value of $L_1$ and $\beta$, we determined the
coefficients of the series
\eq
h=b_1 M+b_3 M^3+b_5 M^5 \dots
\en
which we truncated at order $M^{15}$.
Using the numerical results of $M$ for 8 values of $h$, we computed
the coefficients $b_1$, $b_3$, ..., $b_{15}$.
The optimal choice for these eight values of $h$ is the one for which
 the errors due to the truncation of the series and those
due to  numerical rounding are of the same magnitude. This optimal range
 changes as a function
of $\beta$ and $L_1$. For instance, just to give an idea of the
magnitude of the magnetic fields which we studied, for $\beta=0.36$
and $L_1=24$ the best choice is $h=0.002\, j$ with $j=1\dots 8$~.
The accuracy of the $b_i$ obtained in
this way is decreasing with increasing order. For example, for $\beta=0.37$,
we obtain $b_1$ with 14 significant digits and, e.g., $b_{11}$ with only
3 significant digits.

In principle we could have also used the expansion
\eq
M= a_1 h+ a_3 h^3 + a_5 h^5 \dots \;\;.
\label{1.1}
\en
However, in practice, the use of (\ref{1.1})
leads to results with significantly larger truncation errors,
since in contrast to what happens for $b_i$,
the sign of the coefficients $a_i$ alternates.

Then, for each $\beta$ and $L_1$,
we constructed the $\chi_{2n}$
functions out of the $b_i$ constants. These are the inputs of the FSS analysis
discussed in the next section.
 The relations between the $b_i$ and the
$\chi_{2n}$ can be easily obtained by a straightforward calculation. We report
them here for completeness:
\eqa
\chi_4 &=& -3!\, \frac{b_3}{b_1^4}~,
\label{defchi4}
\\
\chi_6 &=&
- 5!\, \left(\frac{b_5}{b_1^6}-3\frac{b_3^2}{b_1^7}\right)~,
\label{defchi6}
\\
\chi_8 &=&
-7!\, \left(\frac{b_7}{b_1^8}+12\frac{b_3^3}{b_1^{10}}-8\frac{b_3b_5}{b_1^9}
   \right)~,
\label{defchi8}
\\
\chi_{10} &=&
- 9!\, \left(\frac{b_9}{b_1^{10}}-55\frac{b_3^4}{b_1^{13}}-
           10\frac{b_3b_7}{b_1^{11}}
          +55\frac{b_3^2b_5}{b_1^{12}}-5\frac{b_5^2}{b_1^{11}}\right)~,
\label{defchi10}
\\
\chi_{12} &=&
- 11!\, \left({\frac{b_{11}}{{b_1^{12}}}} +
   {\frac{273\,{b_3^5}}{{b_1^{16}}}} -
   {\frac{364\,{b_3^3}\,b_5}{{b_1^{15}}}} +
   {\frac{78\,b_3\,{b_5^2}}{{b_1^{14}}}} +
   {\frac{78\,{b_3^2}\,b_7}{{b_1^{14}}}} -
   {\frac{12\,b_5\,b_7}{{b_1^{13}}}} -
   {\frac{12\,b_3\,b_9}{{b_1^{13}}}} 
   \right). \nonumber \\[-1mm]
\label{defchi12}
\ena
This last step
is the only difference in the present calculation with respect
to the analogous one reported in~\cite{CHPV-00} where we directly studied the
thermodynamic limit of the $b_i$ coefficients. The reason is that as $n$
increases the $\chi_{2n}$ show a much better and smoother FSS behaviour than
the $b_i$.

\subsubsection{The thermodynamic limit}
\label{therlim}

In the second step of the analysis we extrapolate our results to the
thermodynamic limit. We know on general grounds that, since $L_1 \gg \xi$,
the convergence towards the thermodynamic limit of the $\chi_{2n}$ functions is
 exponentially fast, the decay rates being
related to the spectrum of the theory. This result holds for any lattice model.
However, in the case of the two
dimensional Ising model, thanks to the fact that the model can be solved
exactly also on finite lattices~\cite{kau}
(or, in the language of S-matrix theory, thanks
to the fact that the S-matrix of the model is very simple), much more
informations can be obtained on the FSS properties of the free energy and of
its derivatives. In particular, one can explicitly compute
the functional form of the FSS behaviour of
$\chi_{2n}$ (see App. \ref{appb}). It turns out to be
\eq
\chi_{2n}(L_1)-\chi_{2n}(\infty)=L_1^{2n-3/2}
(g_{2n,1}(1/L_1)e^{-mL_1}+g_{2n,2}(1/L_1)e^{-2mL_1}
+g_{2n,3}(1/L_1)e^{-3mL_1}+...)
\label{ne1}
\en
where  $g_{2n,i}(1/L_1)$ are functions
of $1/L_1$, which can be expanded in (positive) powers of $1/L_1$.

By fitting our values of $\chi_{2n}(L_1)$ with this law (by expanding the
functions $g_{2n,i}(L_1)$ and using as free parameters the coefficients of the
first few terms of the expansion) it is possible to
obtain very precise estimates for the thermodynamic limit
$\chi_{2n}(\infty)$. However, this is not the most efficient strategy, since
much of the information contained in the $\chi_{2n}(L_1)$ is lost in the
determination of the coefficients of the Taylor expansion of the
$g_{2n,i}(L_1)$ functions.
Thus, we have elaborated an alternative iterative
procedure which is much simpler to perform than the above multiple fits and
allows to reach
higher precision (up to one additional accurate digit).
The idea behind this Iterative Algorithm (IA, in the following) is first to
absorb the pre-exponential factors of Eq. (\ref{ne1}) in the masses by allowing
them to depend on $L_1$. Then, the resulting
exponential corrections are eliminated by iteratively solving the system of
equations
\begin{eqnarray}
\label{eq1}
\chi_{2n}^{(i)}(L_1-2) &=& c \; \exp(-x (L_1-2)) \;+\; \chi_{2n}^{(i+1)}(L_1)
 \nonumber \\
\chi_{2n}^{(i)}(L_1-1) &=& c \; \exp(-x (L_1-1)) \;+\; \chi_{2n}^{(i+1)}(L_1)
 \nonumber \\
\chi_{2n}^{(i)}(L_1) \phantom{-0)}  &=& c
\; \exp(-x L_1) \phantom{(-2x)} \;+\; \chi_{2n}^{(i+1)}(L_1) \;\;\;
\end{eqnarray}
with respect to  $\chi_{2n}^{(i+1)}(L_1)$, $c$ and $x$.

The index $i$ in $\chi_{2n}^{(i)}$ denotes the order of the iteration, in
particular the $\chi_{2n}^{(0)}$ are the input data of the algorithm
obtained as discussed in the previous section. As $i$ increases the $L_1$
dependence of the $\chi_{2n}^{(i)}$ becomes smaller. After a certain number 
of steps the extrapolation becomes unstable since rounding errors accumulate.
Typically, the
final estimate was obtained by iterating 1, 2, 3 or  4 times, depending on
$\beta$  and on $n$.
The residual
dependence on $L_1$ is used to estimate the error in the best estimate
of $\chi_{2n}(\infty)$.
The results obtained in this way are in perfect agreement with those
obtained by directly fitting Eq. (\ref{ne1}) but, as mentioned above, turn out
to be more precise.
By using the IA  we were able to obtain
reliable estimates of $\chi_n$ for $n\le12$ up to $\beta=0.37$. Our results
are summarized in Tables \ref{chin246} and \ref{chin81012}.

In order  to verify if our method of estimating the errors with the IA method
(quoted in
Tables \ref{chin246} and \ref{chin81012})
is reliable we made the following test.
We studied with the IA a sample of data at a very low value of
 $\beta$ where
 rather large values of $L_1/\xi$ could be reached,
so that the results for the  $\chi_{2n}(L_1)$
on our largest lattices were already good estimates of $\chi_{2n}(\infty)$
with no need of further manipulations.
Then, we performed our analysis using only the data at small $L_1$. In this way
we could explicitly check that our extrapolation scheme gives accurate
results for the thermodynamic limit starting from data such that
$L_1/\xi \approx 7$, which is what we reached
at our largest value of $\beta$, $\beta=0.37$.
Moreover in App.~\ref{testia} we report
a further test of IA.
We applied it to a test function of the type
(\ref{ne1}), finding again support to the reliability of the 
method.

In order to give a feeling of the performances of the IA
for $\beta = 0.37$---this is the value of $\beta$ that is
nearer to the critical point among those we studied and
hence the worst one from the point of view of FSS---we have reported in
Table \ref{iatab} the results of the first four iterations for $\chi_4$ at
$\beta=0.37$. From these numbers we estimate the thermodynamic limit of
$\chi_4$ to be $\chi_4(\infty)=-4.052238(1)$ to be compared with the result
obtained by directly fitting Eq. (\ref{ne1}) which is
$\chi_4(\infty)=-4.052242(10)$.

This same IA was also used in our previous paper~\cite{CHPV-00}, the only
difference being that in that case it was applied to extract the thermodynamic
limit of the $b_i$ constants.
Let us stress, as a final remark,
 that this algorithm is very general and can be used
even when no exact information is available on the FSS behaviour of the
quantity of interest except from the fact that it is dominated by an
exponential decay.

\begin{table}
\begin{center}
\begin{tabular}{|l|l|l|}
\hline
\multicolumn{1}{|c|}{$\beta$} &\multicolumn{1}{c|}{4}
 & \multicolumn{1}{c|}{6}  \\
\hline
0.20  & $-$3.2111149829(1) & \phantom{0}61.94504968(2) \\
0.25  & $-$3.4721394468(1) & \phantom{0}74.5437488(1) \\
0.28  & $-$3.622540033(1)  & \phantom{0}82.255117(3)  \\
0.30  & $-$3.720514859(2)  & \phantom{0}87.459547(5)  \\
0.31  & $-$3.768883189(5)  & \phantom{0}90.081633(5)  \\
0.32  & $-$3.81687386(2)   & \phantom{0}92.71755(1)   \\
0.33  & $-$3.86451567(5)   & \phantom{0}95.36788(5)  \\
0.335 & $-$3.88821530(5)   & \phantom{0}96.6986(1)  \\
0.34  & $-$3.91183945(10)  & \phantom{0}98.0332(1)  \\
0.345 & $-$3.9353922(1)    & \phantom{0}99.3718(1) \\
0.35  & $-$3.9588780(5)    &           100.7142(2) \\
0.355 & $-$3.9823005(10)   &           102.0610(2) \\
0.36  & $-$4.0056653(10)   &           103.4119(2) \\
0.365 & $-$4.028975(1)     &           104.7673(2) \\
0.37  & $-$4.052238(1)     &           106.127(1) \\
\hline
\end{tabular}
\end{center}
\caption{\sl
Thermodynamic-limit results for $\chi_n(t)\, t^{15 n/8 - 2}$, for $n=4,6$.
The quoted error bars are estimates of the systematic error of the
extrapolation.
}
\label{chin246}
\end{table}

\begin{table}
\begin{center}
\begin{tabular}{|l|l|l|l|}
\hline
\multicolumn{1}{|c|}{$\beta$} &\multicolumn{1}{c|}{8}   & \multicolumn{1}{c|}{10}
& \multicolumn{1}{c|}{12} \\
\hline
0.20   & $-$2985.5224(2)  &267883.(1.)   & \phantom{0}$-$23211300.(1000.)\\
0.25   & $-$4002.0604(2)  &400060.(1.)   & \phantom{0}$-$43255500.(1000.) \\
0.28   & $-$4672.309(2)   &494203.(5.)   & \phantom{0}$-$59550000.(10000.) \\
0.30   & $-$5144.20(2)    &563455.(10.)  & \phantom{0}$-$72400000.(10000.) \\
0.31   & $-$5387.715(10)  &600090.(20.)  & \phantom{0}$-$79445000.(30000.) \\
0.32   & $-$5636.31(2)    &638110.(50.)  & \phantom{0}$-$86910000.(100000.) \\
0.33   & $-$5890.10(5)    &677600.(100.) & \phantom{0}$-$94830000.(100000.) \\
0.335  & $-$6018.9(1)     &697800.(100.) & \phantom{0}$-$98950000.(200000.) \\
0.34   & $-$6149.0(1)     &718300.(100.) & $-$103190000.(200000.) \\
0.345  & $-$6280.5(2)     &739000.(500.) & $-$107560000.(200000.) \\
0.35   & $-$6413.1(3)     &760600.(200.) & $-$112000000.(500000.) \\
0.355  & $-$6547.1(3)     &782400.(200.) & $-$116600000.(500000.) \\
0.36   & $-$6682.3(5)     &804400.(300.) & $-$121300000.(1000000.) \\
0.365  & $-$6819.3(3)     &827100.(500.) & $-$126100000.(1000000.) \\
0.37   & $-$6958.1(5)     &850000.(500.) & $-$131000000.(1000000.) \\
\hline
\end{tabular}
\end{center}
\caption{\sl
Thermodynamic-limit results for $\chi_n(t)\, t^{15 n/8 - 2}$ for
$n=8,10,12$. The quoted
error bars are estimates of the systematic error of the extrapolation.}
\label{chin81012}
\end{table}

\begin{table}
\begin{center}
\begin{tabular}{|c|c|c|c|c|c|}
\hline
$L_1$ &$F_4^{(0)}$ &$F_4^{(1)}$ & $F_4^{(2)}$& $F_4^{(3)}$ & $F_4^{(4)}$  \\
\hline
    15 & $-$3.72609418989519 &
 $-$4.07455655     &&      & \\
    16 &  $-$3.80380147054199&
  $-$4.06509498      &&   &  \\
    17 &  $-$3.86341299292473&
  $-$4.05978684  & $-$4.05300296  &   &  \\
    18 &  $-$3.90897624823899 &
  $-$4.05675321  & $-$4.05270713  &   &  \\
    19 &  $-$3.94370300243699 &
  $-$4.05498869  & $-$4.05253535  &  $-$4.05229750  & \\
    20 &   $-$3.97011104914228 &
 $-$4.05394438  & $-$4.05243012  & $-$4.05226371  &  \\
    21 &  $-$3.99015681870086 &
  $-$4.05331548  & $-$4.05236336  & $-$4.05224754  &  $-$4.05223270 \\
    22 &   $-$4.00535064954957 &
 $-$4.05293008  & $-$4.05232014  & $-$4.05224075  &  $-$4.05223585 \\
    23 &  $-$4.01685278847850 &
  $-$4.05268980  & $-$4.05229191  & $-$4.05223874  & $-$4.05223789 \\
    24 &   $-$4.02555119351443 &
 $-$4.05253744  & $-$4.05227342  & $-$4.05223837    & $-$4.05223829 \\
\hline
\end{tabular}
\end{center}
\caption{\sl Results of the IA analysis for $F_4\equiv \chi_4 t^{-11/2}$
at $\beta=0.37$.}
\label{iatab}
\end{table}

\subsection{Small-$t$ expansion of $\chi_n(t)$}
\label{smalltexp}

As it is well known,
the TM approach gives reliable results only for rather small
values of $\beta$. In order to perform the extrapolation $\beta\to\beta_c$,
it is thus mandatory to have a good control of the scaling corrections of
$\chi_n(t)$.
In this section we address this problem in detail. Our goal will be to obtain
for each $\chi_{2n}$ the exact form of the scaling function up to $O(t^4)$ and
the spectrum of the possible scaling dimension (i.e. the exponents of the
corresponding contributions in the scaling function) up to $O(t^5)$).

The most important information which is needed to perform this analysis is the
spectrum of the irrelevant operators of the theory. This problem was recently
addressed numerically in Refs. \cite{GMC-88,Nickel,Guttmann_private} where it
was shown that, for rotationally invariant quantities like the
free energy and its derivatives, the first correction due to the
irrelevant operators in the
square-lattice Ising model appears at order $t^4$ (hence  no correction
of order $t^2$ is present) and that the next
irrelevant operator contributes only at order $t^6$.
This result has been shown for the susceptibility
in zero field,
but a standard application of RG ideas implies that it should also apply
to the free energy as a function of $t$ and $h$.\footnote{This pattern agrees
with an independent analysis, in the framework of CFT, based
on the spectrum of the quasi-primary operators of the critical Ising model.
This result was already anticipated in Ref.~\cite{CHPV-00} and will be
discussed in full detail in a forthcoming publication.} This is also
confirmed by the less precise numerical results of Ref. \cite{ch99}
for the free energy on the critical isotherm and by the analytic study
of the two-point function at large distances \cite{cccpv}.

 This result allows us to determine the scaling corrections to $\chi_{2n}(t)$
 up to order $t^4$ by using standard RG techniques
 (see \cite{af,ssv1,CHPV-00}).

As a first step, let us write the free energy of the model
in terms of nonlinear scaling fields \cite{Wegner_1976}:
\begin{eqnarray}
F(t,h) & = & F_{b}(t,h) +
   |u_t|^{d/y_t} f_{\rm sing}\left(\frac{u_h}{|u_t|^{y_h/y_t}},
       \left\{\frac{u_j}{|u_t|^{y_j/y_t}} \right\}\right)
\nonumber \\
&& +
   |u_t|^{d/y_t} \log |u_t|
      \widetilde{f}_{\rm sing}\left(\frac{u_h}{|u_t|^{y_h/y_t}},
       \left\{\frac{u_j}{|u_t|^{y_j/y_t}} \right\}\right).
\label{scal0}
\end{eqnarray}
Here $F_{b}(t,h)$ is a regular function of $t$ and $h^2$,
$u_t$, $u_h$, $\{u_j\}$ are the
nonlinear scaling fields associated respectively to the
temperature, the magnetic field and the irrelevant operators, and
$y_t$, $y_h$, $\{y_j\}$ are the corresponding dimensions.
For the Ising model $y_t = 1$ and $y_h= 15/8$.
Notice the
presence of the logarithmic term, that is related to a ``resonance"
between the thermal and the identity operator.\footnote{In principle,
other logarithmic terms may arise from additional
resonances due to the fact
that $y_j$ are integers or differ by integers from $y_h$,
and indeed they have been observed in a high-precision analysis
of the asymptotic behaviour of the susceptibility
\cite{Guttmann_private}. They will not
be considered here since these contributions either are subleading with
respect to those we are interested in or have a form that is already included.}
Since all
numerical data indicate that for $t\to0$ all zero-momentum correlation
functions diverge as a power of $t$ without logarithms---our results
provide additional evidence for this cancellation---we shall
assume in the following (as in Ref. \cite{af})
that $\widetilde{f}_{\rm sing}$ does not depend on its first argument
${u_h}/{|u_t|^{y_h/y_t}}$. There is also some evidence that the
leading contribution due to the irrelevant operators is absent.
Indeed, for the susceptibility one would expect a correction of order
$t^4\log |t|$ which is not found in the high-precision study of the
susceptibility reported in \cite{Guttmann_private}. In the following we will be
conservative and we will report results with and without corrections
of order $t^4\log |t|$. Our final estimates assume however that such a term
is absent.

The scaling fields are analytic functions of $t$ and $h$ that
respect the ${\bf Z_2}$ parity of $h$. Let us write their
 Taylor expansion, keeping only those terms
that are needed for our analysis (we use the notations of Ref.~\cite{af}):
\eq
u_h~=~h~[1~+~c_ht~+~d_ht^2~+~e_h h^2~+~ f_h t^3 ~+~O(t^4,th^2)],
\label{u_h}
\en
\eq
u_t~=~ t ~+~b_t h^2 ~+~c_tt^2 ~+~d_t t^3 ~+~e_tth^2 ~+~
~ f_t t^4 ~+ O(t^5,t^2h^2,h^4).
\label{u_t}
\en
All these coefficients (except $e_h$)
have been determined exactly or numerically with very high
precision, see \cite{ssv1,Nickel,CHPV-00}. We list them here for completeness:
\eq
c_h=\frac{\beta_c}{\sqrt{2}},  \hskip1cm
d_h=\frac{23 \beta_c^2}{16},   \hskip1cm
f_h=\frac{191 \beta_c^3}{48\sqrt{2}},
\label{eq24}
\en
\eq
c_t=\frac{\beta_c}{\sqrt{2}},\hskip1cm
d_t=\frac{7 \beta_c^2}{6}, \hskip1cm
f_t=\frac{17 \beta_c^3}{6\sqrt{2}},
\label{eq25}
\en
\eq
e_t = b_t \beta_c \sqrt{2}, \qquad \qquad b_t = - {E_0 \pi\over 16 \beta_c^2},
\en
where \cite{Kong-unpublished}
\eq
      E_0=0.0403255003...
\label{E0}
\en
is the coefficient of the contribution proportional
to $t\log |t|$ in the susceptibility.

The coefficient $e_h$ is the only term that was not reported in \cite{CHPV-00}.
In App. \ref{appc} we show that it is possible to determine it
by analyzing the scaling corrections to
the free energy on the critical isotherm. Using the precise
data of Ref. \cite{ch99}, we obtain
\eq
      e_h= -0.00727(15).
\en
Using Eq. (\ref{scal0}) and the expansions of the scaling fields reported
above, we can compute the leading terms in the asymptotic
expansion of $\chi_n(t)$ for $t\to 0$.
We obtain
\begin{equation}
\chi_n(t) =\, t^{2 - 15 n/8} a_{n}(t) +
              t^{19/4 - 15 n/8} b_{n}(t) + ... \;,
\label{scalchi}
\end{equation}
where
\begin{eqnarray}
a_{n}(t) &=& C^+_n (1 + \alpha_1 \beta_c t +
    \alpha_2 \beta_c^2 t^2 + \alpha_3 \beta_c^3 t^3 + O(t^4) ),
\\
b_{n}(t) &=& C_{n-2}^+ (\zeta_0 + \zeta_1 t + O(t^2)).
\end{eqnarray}
The coefficients $\alpha_i$ and $\zeta_i$ are given by:
\begin{eqnarray}
\alpha_1 &=& - {7 n - 16\over 8 \sqrt{2}}, \\
\alpha_2 &=& {147 n^2 - 1080 n + 2176\over 768}, \\
\alpha_3 &=& - {343 n^3 - 5208 n^2 + 26240 n -49152\over 6144 \sqrt{2}}, \\
\zeta_0 &= & {E_0 n (n - 1) (15 n - 46) \pi\over 128 \beta_c^2}, \\
\zeta_1 & =&  - {E_0 n (n - 1) (7 n - 38) (15 n - 46) \pi \over
            1024 \sqrt{2} \beta_c} +
          n (n - 2) (n - 1) e_h.
\end{eqnarray}
Plugging these coefficients into Eq. (\ref{scalchi}) we obtain the exact form of
the scaling function up to the contribution of the first irrelevant field, i.e.
up to $O(t^4)$. The simplest way to use the exact knowledge of these terms of
the scaling function to extract the amplitudes $C^+_n$
 is to construct the quantity:
\eq
M^{(n)}(t)\equiv \frac{\chi_n(t)\, t^{15n/8-2}-
                       b_{n}(t)t^{15n/8-19/4}}{a_{n}(t)}
\label{scal1}
\en
which has the following expansion for $t\to 0$:
\eq
M^{(n)}(t) = C^+_n (1+p_1 t^4 +\tilde p_1 t^4~\log t  +p_2 t^{4.75}
                  + O(t^5)).
\label{scal2}
\en
The terms proportional to $t^4$ and to $t^{4.75}$ are due to
 the first unknown coefficients of $a_{n}(t)$ and $b_{n}(t)$ respectively.
They also take into account the possible presence of irrelevant operators
contributing to order $t^4$.
The term proportional to $t^4~\log t $ is the only remnant of the
 $\widetilde{f}_{\rm sing}$ term in Eq. (\ref{scal0}) and it is due
 to the irrelevant operators in $\widetilde{f}_{\rm sing}$.

The constants $C^+_n$
  are determined in sequence. We start by using $C_2^+$,
  which is known to very high precision,
  $C_2^+ = 0.9625817323087721140443\ldots$
  \cite{mccoy,Nickel,Guttmann_private},
  to estimate $C_4^+$ with a best-fit analysis of Eq.
  (\ref{scal2}) with $n=4$.
 The value of $C_4^+$ obtained in this way
 is then used as input to construct the function $b_{6}(t)$ which appears in
 Eq. (\ref{scal1}), thus allowing to estimate $M^{(6)}(t)$. At this
point, by using again Eq. (\ref{scal2}), we obtain the best-fit value for
 $C_6^+$ and repeat the whole procedure for the next value of $n$.

As a consequence, in the determination of $M^{(n)}(t)$
there are three sources of uncertainty:
\begin{description}
\item{a]} The uncertainty in the TM estimates of $\chi_n(t)$ which we use
as input of our analysis.
\item{b]} The uncertainty in the estimate of $e_h$ and $E_0$.
\item{c]} The uncertainty in our best estimate of $C^+_{n-2}$.
\end{description}

These uncertainties must be treated in different ways. [a] can be
straightforwardly propagated to $M^{(n)}(t)$.
On the contrary [b] and [c] appear as uncertainties of
the whole fitting function.
To deal with them we followed
 the simplest (and most conservative) strategy.
 Let us study as an example
 the uncertainty due to the error on $e_h$.
We constructed two sets of data, $M_+^{(n)}(t)$ and $M_-^{(n)}(t)$,
obtained by using $e_h+\delta e_h$ and
$e_h-\delta e_h$ respectively, and then, for each set,
we determined the best-fit values of the parameters.
 The difference between these results is a conservative estimate of
the error induced by the uncertainty on $e_h$.
In order to give an idea of the size of these uncertainties, we
have reported in Table \ref{fit2}
those induced by $e_h$ in the best-fit estimates of $C^+_6$.

It turns out that the errors
induced by the uncertainties on
$C^+_{n-2}$ and $E_0$ are always negligible with respect to that due to
 $e_h$
and, what is more important,
that this last one is in any case negligible with respect to the
uncertainty in the best-fit estimate,
i.e. with respect to the uncertainty due to the lack of
knowledge of the higher-order correction terms. For this reason
 we shall neglect the errors of type [b] and [c] in our
 final results.

\subsection{The fitting procedure}

To analyze the data we followed a procedure similar to that presented in
Ref.~\cite{CHPV-00}.
For each set of data we performed five different fits of $M^{(n)}$
using Eq. (\ref{scal2}),
\begin{description}
\item{f1]} keeping $C^+_n$ and $p_1$ as free parameters, and setting
           $\tilde p_1 = p_2=0$;
\item{f2]} keeping $C^+_n$ and $\tilde p_1$ as free parameters, and setting
           $p_1=p_2=0$;
\item{f3]} keeping $C^+_n$, $p_1$, and $\tilde p_1$ as free parameters,
           and setting $p_2=0$;
\item{f4]} keeping $C^+_n$, $p_1$, and $p_2$ as free parameters,
           and setting $\tilde p_1 = 0$;
\item{f5]} keeping $C^+_n$, $p_1$, $\tilde p_1$, and $p_2$
           as free parameters.
\end{description}

In order to estimate the systematic errors involved in the determination of
$C_n^+$,
we performed for all the fitting functions several independent fits,
trying first to fit all the existing data (reported in Tables \ref{chin246} and
\ref{chin81012})
and then eliminating the data one by one,
starting from the farthest from the critical
point. Among the
set of estimates of the critical amplitudes we selected only those
fulfilling the following requirements:

\begin{description}
\item{1]} \phantom{X}
The reduced $\chi^2$ of the fit must be of order unity. In order to fix
precisely a threshold, we required the fit to have a confidence level larger
than $30\%$.
\item{2]} \phantom{X}
For all the subleading terms included in the fitting function, the amplitude
estimated from the fit must be larger than the corresponding error.
\item{3]} \phantom{X}
If the fit contains $k$ free parameters besides $C_n^+$, then at least
$2^k$ degrees of freedom must be used in the fit. This means that in fits of
type f1 and f2 at least four data must be used ($k=1$: two parameters plus two
 degrees of freedom) in those of type f3 and f4 at least 7 data must be used
($k=2$: three parameters plus four degrees of freedom). In f5 at least 12 data
must be used ($k=3$: four parameters plus eight degrees of freedom)
\end{description}

This last constraint is a
generalization to higher values of $n$ of the  one that we proposed
in~\cite{CHPV-00}. It seems to encode very well
the real statistical significance of the data.

Finally, among all the
estimates of the critical amplitude
$C_n^+$ fulfilling these requirements
we selected the smallest and the largest
ones as lower and upper bounds. We consider the mean of these two values as
our best prediction for $C_n^+$. The values are reported in
Table~\ref{fit1}. The errors quoted in Table~\ref{fit1} are half of the
difference between the upper and lower bounds. They seem to
give a reliable estimate of
the uncertainty of our results.

\begin{table}[tbp]
\vskip 0.2cm
\begin{center}
\begin{tabular}{|l|l|l|}
\hline
     &  $\tilde p_1\not= 0$ & $\tilde p_1=0$     \\
\hline
$C_4^+$    & $-$4.379095(8) & $-$4.379094(6)   \\
$C_6^+$    & 125.9330(11) &  125.9332(6)   \\
$C_8^+$    & $-$9066.5(9) & $-$9066.4(7)   \\
$C_{10}^+$    & $1216.33(80)\times 10^3$ & $1216.34(60)\times 10^3 $   \\
$C_{12}^+$    & $-26260(150)\times 10^4 $& $ -26175(60)\times 10^4 $   \\
\hline
\end{tabular}
\end{center}
\vskip 0.2cm
\caption{\sl Best-fit estimates of the amplitudes $C_n^+$.
}
\label{fit1}
\end{table}

Let us briefly comment on these results:
\begin{description}
\item{a]}
In Table \ref{fit1} we have reported in two separate columns the
best-fit results
with and without the log-type contribution $\tilde p_1 t^4~\log t $. The reason
of this choice is that there are strong numerical
indications~\cite{Guttmann_private} that
$\tilde p_1 =0$ in the Ising model. To keep our analysis as general as possible,
we report also the result with $\tilde p_1 \not= 0$.
It is  interesting to note that
the estimates of $C_n^+$ ($n\leq 10$) do not depend essentially on this choice,
while the error decreases by a factor 1.5--2 if we assume $\tilde p_1 =0$.
This seems to indicate
that the fitting procedure is stable and reliable.

\item{b]}
In the first line of Table \ref{fit1} we report our best-fit estimate of
$C_4^+$, which we had already estimated
in~\cite{CHPV-00} with the same techniques used in the present paper.
 The only improvement
with respect to \cite{CHPV-00} is that we now also know exactly the value
of $e_h$ which was the dominant (of order $t^{3.75}$)
unknown correction in~\cite{CHPV-00}.
It is instructive to compare the two estimates.
In \cite{CHPV-00} we obtained, keeping also into account the
log-type corrections,
$C_4^+=-4.379101(9)$, while the present value is
$C_4^+=-4.379095(8)$. The two results are, as expected, fully
compatible. The new one is only slightly more precise than the previous one.
The fact that the enhancement in precision is so small
is due to the fact that, by using the exact value of $e_h$,
we only improve the known small-$t$ expansion of $\chi_n(t)$
from $O(t^{3.75})$ to  $O(t^{4})$.

\item{c]} It is instructive to look in more detail to the results of the fits
in one particular case. Let us consider, for instance, $C_6^+$.
The results are reported in Table \ref{fit2}, where, for each type of fit,
we quote the two fits which correspond to
the highest and lowest values of $C_6^+$
(in quoting the uncertainties in the
best-fit estimates we also report those induced by $e_h$, to give an idea of
their size).

\begin{table}[tbp]
\vskip 0.2cm
\begin{center}
\begin{tabular}{|l|l|l|}
\hline
$C_6^+$     &    d.o.f. &  fit type     \\
\hline
$125.93274(12)(0)$    & 2  & f1 \\
$125.93298(5)(0)$    & 9  & f1 \\
\hline
$125.93301(14)(2)$    & 2  & f2 \\
$125.93389(16)(7)$    & 7  & f2 \\
\hline
$125.93223(30)(1)$    & 5  & f3 \\
$125.93341(15)(2)$    & 8  & f3 \\
\hline
$125.93290(14)(0)$    & 7  & f4 \\
$125.93369(12)(1)$    & 9  & f4 \\
\hline
$125.93253(20)(2)$    & 8  & f5 \\
$125.93258(7)(2)$    & 10  & f5 \\
\hline
\end{tabular}
\end{center}
\vskip 0.2cm
\caption{\sl
Estimates of $C^+_6$.
 In the first column we report the best-fit results for the critical amplitude
(the first  error  in parenthesis is that induced by the systematic errors
of the input data while the second is due to the error on $e_h$),
in the second column
the number of degrees of freedom (i.e. the number of data used in the fit minus
the number of free parameters) and in the last column the type of fit.}
\label{fit2}
\end{table}

The most impressive feature of this set of fits is
 that with only the $t^4$ contribution
 besides
$C_6^+$ (fit f1) one can fit all the data up to $\beta=0.31$.

Looking at Table \ref{fit2}, we obtain our best estimate
\eq
C_6^+=125.9330(11).
\label{C6-est1}
\en
If we additionally assume that $\tilde p_1=0$,
i.e. we keep into account only the results of
the fits of type f1 and f4, then we find
\eq
C_6^+=125.9332(6).
\label{C6-est2}
\en
The estimates (\ref{C6-est1}) and (\ref{C6-est2}) are those reported in
Table \ref{fit1}.

\end{description}

\subsection{Summary of the results}

Using the estimates of $C^+_{2n}$ obtained in the previous section,
we can estimate $r_{2n}$ by using Eqs. (\ref{defr6})-(\ref{defr12}).
We finally obtain
\begin{eqnarray}
&& g_4 = 14.697323(20),   \label{g4est} \\
&& r_6 = 3.67866(3)(2), \label{r6est} \\
&& r_8 = 26.041(8)(3), \label{r8est} \\
&& r_{10} = 284.5(1.4)(1.0),  \label{r10est} \\
&& r_{12} = 4200(320)(420).  \label{r12est}
\end{eqnarray}
For each $r_{2n}$ we report two errors: the first one is due to the
error on $C^+_{2n}$, while the second one expresses the uncertainty
due to the error on all $C_{2k}^+$, $k<n$.
Note that the high precision reached in
the estimates of $C_{2n}^+$ is partially lost in the estimates of
$r_{2n}$ because of the cancellations among the various terms in
the sums.
This effect is particularly important for $r_{10}$ and $r_{12}$.

\begin{table}[tbp]
\vskip 0.2cm
\begin{center}
\begin{tabular}{|l|l|l|}
\hline
        Method &  Ref. & $g_4$     \\
\hline
         TM+CFT & [this work] & 14.697323(20)   \\
         FF & \cite{b2000} & 14.6975(1)   \\
         HT & \cite{PV-gre} & 14.694(2)   \\
         HT & \cite{Butera-Comi} & 14.693(4)   \\
         HT & \cite{ZLF} & 14.700(17)\\
         MC & \cite{b2000} & 14.69(2)   \\
         MC & \cite{Kim} & 14.7(2)   \\
         FT $\epsilon$-expansion & \cite{PV-gre-2} & 14.7(4) \\
         FT $g$-expansion & \cite{LZ-77} & 15.5(8) \\
         FT $g$-expansion & \cite{OS-00} & 15.4(3) \\
         $d$-expansion & \cite{BB-92} & 14.88(17) \\
\hline
\end{tabular}
\end{center}
\vskip 0.2cm
\caption{\sl
Estimates of $g_4$. We also report the
existing results from the form-factor approach (FF),
high-temperature expansions (HT), Monte Carlo simulations (MC),
field theory (FT) based on the $\epsilon$-expansion
and the fixed-dimension $d=2$ $g$-expansion,
and a method based on a dimensional expansion around $d=0$
($d$-expansion).
}
\label{sumgr}
\end{table}

\begin{table}[h]
\vskip 0.2cm
\begin{center}
\begin{tabular}{|l|l|l|l|l|l|l|}
\hline
&TM+CFT  & HT \cite{PV-effpot} & HT \cite{ZLF} & MC \cite{Kim} &
  FT $\epsilon$-exp. \cite{PV-effpot} &FT $g$-exp. \cite{SO-98} \\
\hline
$r_6$    & 3.67866(5) & 3.678(2) & 3.679(8) & 3.93(12) & 3.69(4) & 3.68 \\
$r_8$    & 26.041(11) & 26.0(2) & &    28.0(1.6) & 26.4(1.0) & \\
$r_{10}$ & 284.5(2.4) & 275(15) &  & & &  \\
$r_{12}$ & 4200(740)  &  & & & &\\
\hline
\end{tabular}
\end{center}
\vskip 0.2cm
\caption{\sl
Estimates of $r_{2n}$. We also report the
existing results from
high-temperature expansions (HT), Monte Carlo simulations (MC), and
field theory (FT) based on the $\epsilon$-expansion
and the fixed-dimension $d=2$ $g$-expansion.
}
\label{sumrj}
\end{table}

Tables \ref{sumgr} and \ref{sumrj} compare
our results with the existing  data from other approaches.
Our estimates of $r_{2n}$
perfectly agree with those obtained analyzing HT expansions.
On the contrary, the Monte Carlo results for $r_{2n}$ of
Ref. \cite{Kim} are systematically larger. This could be an indication
that the finite-size scaling curves obtained in Ref. \cite{Kim}
are still affected by large scaling corrections.

\section{The critical equation of state}
\label{sece3}

\subsection{General features}

The basic result  of the RG theory is that asymptotically
close to the critical point the equation of state may be written in the scaling form \cite{Griffiths-67}
\begin{equation}
h = {\partial{\cal F}\over \partial M} \propto
    M |M|^{\delta-1} f(x), \qquad\qquad x \propto t |M|^{-1/\beta},
\label{eqstfx}
\end{equation}
where $f(x)$ is a universal scaling function
normalized in such a way that $f(-1)=0$ and $f(0)=1$. The value
$x=-1$ corresponds to the coexistence curve, and $x=0$ to
the critical point $t=0$.
The function $h(M,t)$, representing the external field in the critical
equation of state, satisfies Griffiths' analyticity, i.e. it is
regular at $t=0$ for $M>0$ fixed and at $M=0$ for $t>0$ fixed.
This implies that $f(x)$ is analytic at $x=0$, and
it has a regular  expansion  for large-$x$ of the form
\begin{equation}
f(x) = x^\gamma \sum_{n=0}^\infty f_n^\infty x^{-2n\beta}.
\label{largexfx}
\end{equation}

As already mentioned in the introduction, many things are exactly
known for the two-dimensional Ising model.
However, there is no exact expression
for the free energy and for the critical equation of state
in the whole $(t,h)$ plane.
In Table \ref{eqstd2} we report a summary of the known results
for the two-dimensional Ising model (there we consider only infinite-volume
quantities).
Many of them are known exactly, for the others we report their best
estimate. The results that have not been derived in this paper
have been taken from Refs.~\cite{WMTB-76,Delfino-98,CPRV-varie,PV-99,CHPV-00}.

\begin{table}
\begin{center}
\begin{tabular}{|l|c|}
\hline
\multicolumn{2}{|c|}{Critical exponents and amplitude ratios} \\
\hline\hline
$\gamma$ &  7/4 \\
$\nu$ &  1 \\
$\eta$ &  1/4 \\
$\beta$ &  1/8 \\
$\delta$ &  15 \\
$\omega$ &  2 \\ \hline
$U_0\equiv A^+/A^-$ & 1  \\
$U_2\equiv C^+_2/C^-_2$ & 37.69365201    \\
$R_c^+\equiv A^+C^+_2/B^2 $ & 0.31856939 \\
$R_c^-\equiv A^- C^-_2/B^2 $ & 0.00845154 \\
$R_\chi\equiv Q_1^{-\delta}\equiv C^+_2 B^{\delta-1}/(\delta C^c)^\delta$
& 6.77828502  \\
$w^2\equiv C^-_2 /[ B^2 (f^-)^2]$ &  0.53152607 \\
$U_\xi\equiv f^+/f^- $ &  3.16249504 \\
$U_{\xi_{\rm gap}}\equiv f^+_{\rm gap}/f^-_{\rm gap}$ & 2 \\
$Q^+ \equiv A^+ (f^+)^2$  &  0.15902704  \\
$Q^- \equiv A^- (f^-)^2 $  &  0.015900517 \\
$Q^+_\xi\equiv f^+_{\rm gap}/f^+$ & 1.000402074 \\
$Q^c_\xi\equiv f^c_{\rm gap}/f^c$&  1.0786828  \\
$Q^-_\xi\equiv f^-_{\rm gap}/f^-$&  1.581883299 \\
$Q_2\equiv (f^c/f^+)^{2-\eta} C^+_2/C^c$ &  2.8355305\\\hline
$g_4\equiv -C_4^+/[ (C^+_2)^2 (f^+)^2] $ & 14.697323(20)  \\
$R_4^+\equiv - C_4^+B^2 / (C^+_2)^{3}$ & 7.336774(10) \\
$r_6\equiv g_6/g_4^2$ & 3.67866(5) \\
$r_8\equiv g_8/g_4^3$ & 26.041(11)\\
$r_{10}\equiv g_{10}/g_4^4$ & 284.5(2.4)\\
$r_{12}\equiv g_{12}/g_4^5$ & $4.44(6) \times 10^3$ \\
$r_{14}\equiv g_{14}/g_4^6$ & $8.43(3) \times 10^5$ \\
$v_3\equiv - C_3^-B/(C^-_2)^2$  & 33.011(6)  \\
$v_4 \equiv -C_4^-B^2/(C^-_2)^3 + 3 v_3^2 $ & 48.6(1.2) \\
\hline
\end{tabular}
\end{center}
\caption{\sl
Critical exponents and universal amplitude ratios for the
two-dimensional Ising universality class. See App. \ref{AppA} for
the definitions. 
}
\label{eqstd2}
\end{table}

In the following we will determine the equation of state, starting
from its expansion for small magnetization in the HT phase.
It is therefore useful to introduce a different representation
that is analytic for $M\to 0$.
Using the results of Secs. \ref{sece1} and \ref{smalltexp}, in particular 
Eq.~(\ref{scal0}) and the discussion following it,
one may write the Helmholtz free energy as
\begin{equation}
{\cal F}_{\rm sing}(t,M) = a t^2 V(z) + {A\over2} t^2 \log |t|
\label{ffff}
\end{equation}
where  
\begin{eqnarray}
&z = b |M| t^{-\beta}, \qquad \qquad & V(z) = {z^2\over 2} + {z^4\over 4!} + O(z^6), \\
&a = - {(C^+)^2/C^+_4},\qquad\qquad  &b = \left[-{C^+_4\over
(C^+)^3}\right]^{1/2}.  \nonumber
\end{eqnarray}
The constant $A$ is related to the amplitudes of the specific
heat for $h\to 0$ defined in Eq.~(\ref{sphamp}). Indeed the
analyticity of the free energy 
for $t=0$, $h \not= 0$ implies $A^+ = A^- \equiv A$,
and thus $U_0\equiv A^+/A^-=1$.
The presence of the logarithmic term gives rise to logarithms in the 
expansions of $V(z)$ for $z\to\infty$. 
Indeed, the analyticity of ${\cal F}_{\rm sing}(t,M)$ for $t=0$, 
$|M|\not=0$ implies, for large $z$, 
\begin{equation}
V(z) = z^{16} \sum_{k=0} c_k z^{-8k} + c_{\rm log} \log z ,
\label{espansioneA1z-Ising2d}
\end{equation}
The constant $c_{\rm log}$ is easily expressed in terms 
of invariant amplitude ratios: 
\begin{equation} 
    c_{\rm log} = {4 A\over a} = 4 Q^+ g_4^+.
\end{equation}
For the equation of state we have 
\begin{equation}
h = {\partial{\cal F}\over \partial {M}} = a \,b \,t^{15/8} {\partial
          V(z)\over\partial z} \equiv 
          a \,b \,t^{15/8} B(z) 
\label{eqa}
\end{equation}
where, using Eq. (\ref{AZ}),
\begin{equation}
B(z) = z + \case{1}{6}z^3 + \sum_{j=3} {r_{2j}\over (2j-1)!} z^{2j-1}.
\label{Bzdef}
\end{equation}
For large $z$, using (\ref{espansioneA1z-Ising2d}) we obtain the expansion
\begin{equation}
B(z) = z^{15} \sum_{k=0} B^{\infty}_k z^{-8k}.
\label{asyFz}
\end{equation}
The constant $B_0^\infty$ can be expressed in terms of invariant
amplitude ratios:
\begin{equation}
B_0^\infty = R_\chi (R_4^+)^{(1-\delta)/2} = 0.592357(6)
\times 10^{-5},
\label{f0inf}
\end{equation}
where we have used the numerical results of Table~\ref{eqstd2}.
Moreover, by using Eq. (\ref{espansioneA1z-Ising2d}),
we obtain
\begin{equation}
B_2^\infty = c_{\rm log} = 4 Q^+ g_4 = 9.349087(13).
\label{F2infty-Ising2d}   
\end{equation}
To reach the coexistence curve, corresponding to \ $t<0$ and $h=0$, one should
perform an analytic continuation in the complex $t$-plane
\cite{ZJbook,GZ-97}.  The spontaneous magnetization is related to
the complex zero $z_0=|z_0|e^{-i\pi\beta}$  of $B(z)$~\cite{GZ-97},
where
\begin{equation}
|z_0|^2 = R_4^+\equiv - C_4^+B^2 / (C^+_2)^{3}  = 7.336774(10).
\label{z0}
\end{equation}
Therefore, the description of the
coexistence curve is related to the behaviour of $B(z)$ in the
neighbourhood of $z_0$.

The functions $B(z)$ and $f(x)$ give equivalent representations
of the equation of state. Indeed, they are simply related by
\begin{equation}
z^{-\delta} B(z) = B_0^\infty f(x), \qquad\qquad z = |z_0| x^{-\beta}.
\label{relazioneF-f}
\end{equation}

\subsection{Parametric representations}

In order to obtain a representation of the
critical equation of state that is valid in the whole critical region,
we need to extend analytically the expansion (\ref{Bzdef})
to the low-temperature region $t<0$. For this purpose,
one may use parametric representations, which implement in a simple
way all scaling and analytic properties \cite{Schofield-69,SLH-69,Josephson-69}.
One may parametrize $M$ and $t$
in terms of $R$ and $\theta$ according to
\begin{eqnarray}
M &=& m_0 R^\beta \theta  ,\label{parrep} \\
t &=& R(1-\theta^2), \nonumber \\
h &=& h_0 R^{\beta\delta}h(\theta),\nonumber
\end{eqnarray}
where $h_0$ and $m_0$ are normalization constants.
The variable $R$ is nonnegative and measures
the distance from the critical point in the $(t,h)$ plane;
the critical behaviour is obtained for $R\to 0$.
The variable $\theta$  parametrizes the displacements along the lines
of constant $R$.
The line $\theta=0$ corresponds to the HT phase $t>0$ and $h=0$,
the line $\theta=1$ to the critical isotherm $t=0$,
and $\theta=\theta_0$, where $\theta_0$ is the smallest positive zero
of $h(\theta)$ to the coexistence curve $T<T_c$ and $h\to 0$.
Of course, one should have $\theta_0 > 1$.
The regularity properties of the critical equation of state
require  $h(\theta)$ to be  analytic in the physical domain
$0\le\theta<\theta_0$.
This is at variance with what happens for the scaling functions
$f(x)$ and $B(z)$, that are nonanalytic for $x\to\infty$ and $z\to \infty$
respectively.
This fact is important from a practical point of view.
Indeed, in order to obtain approximate expressions
of the equation of state, one can approximate $h(\theta)$
with analytic functions. The structure of the parametric
representation automatically ensures the correct analytic properties
of the equation of state.

Note that the  mapping (\ref{parrep}) is invertible only in the region
$\theta<\theta_{l}$ where
\begin{equation}
\theta_{l}^2 = {1\over 1-2\beta}={4\over 3}.
\end{equation}
Thus, the physically relevant interval
$0\leq \theta \leq \theta_0$ must be contained in the region
$\theta<\theta_{l}$, and thus we should have $\theta_0 < \theta_l$.
In practice, since $\theta_l$ is a singular point of the mapping,
it is important that $\theta_l - \theta_0$ is not too small.
As we shall see, all our approximations satisfy this condition.

The function $h(\theta)$ is odd in $\theta$, and
is normalized so that $h(\theta)=\theta+O(\theta^3)$.
Since $M = C^+_2 h t^{-\gamma}$ for $M\to 0$, $t>0$, this condition
implies $m_0 = C^+ h_0$. Following Ref. \cite{GZ-97}, we then replace
$h_0$ by a single normalization constant $\rho$ in such a way that
we can write
\begin{eqnarray}
&&z = \rho \,\theta\,\left( 1 - \theta^2\right)^{-\beta},
\label{thzrel} \\
&&B(z(\theta)) = \rho \left( 1 - \theta^2 \right)^{-\beta\delta} h(\theta).
\label{hFrel}
\end{eqnarray}
In the exact parametric equation the value of $\rho$ may be chosen
arbitrarily: clearly the physical function $B(z)$
does not depend on it. However, if we adopt an approximation for $h(\theta)$,
as we will do, the dependence of $B(z)$ on $\rho$ is not eliminated.
One may then choose $\rho$ to obtain an optimal approximation.

{}From the function $h(\theta)$ one may calculate the scaling
functions $f(x)$, using the relations
\begin{eqnarray}
&& x = {1 - \theta^2\over \theta_0^2 - 1} \left( {\theta\over
\theta_0} \right)^{-1/\beta} ,\label{fxmt}\\
&& f(x) = \theta^{-\delta} {h(\theta)\over h(1)},\nonumber
\end{eqnarray}
and all universal
amplitude ratios involving zero-momentum quantities,
such as the $n$-point susceptibilities
(see, e.g., Ref.~\cite{CPRV-99} for a list of formulae).

\subsection{Polynomial approximations for $h(\theta)$}

In order to construct approximate parametric representations, we consider
a systematic approximation
scheme based on polynomial approximations of $h(\theta)$~\cite{GZ-97}, i.e.
\begin{equation}
h^{(k)}(\rho,\theta) =
  \theta  + \sum_{i=1}^{k-1} h_{2i+1}(\rho) \theta^{2i+1}.
\label{hexpn}
\end{equation}
The coefficients $h_{2i+1}(\rho)$ are functions of $\rho$, $\gamma$, $\beta$,
and are obtained by matching the small-$z$ expansion
of $B(z)$ to $O(z^{2k-1})$, cf. Eq. (\ref{hFrel}).
This kind of approximation turned out to be effective for the
determination of the critical equation
of state of three-dimensional Ising-like systems~\cite{GZ-97,CPRV-99},
and was generalized to models with Goldstone singularities
\cite{CPRV-eqst}.
In order to optimize $\rho$ for a given truncation
$h^{(k)}(\rho,\theta)$, we use a procedure based on the physical
requirement of minimal dependence on $\rho$ of the resulting universal
function
\begin{equation}
B^{(k)}(\rho,z) \equiv  \rho \left( 1 - \theta^2 \right)^{-\beta\delta}
h^{(k)}(\rho,\theta).
\label{ftapp}
\end{equation}
One may indeed prove~\cite{CPRV-99} that
for any truncation $k$ there exists a solution
$\rho_k$ independent of $z$ that satisfies a global stationarity
condition, i.e.
\begin{equation}
\left. \partial B^{(k)}(\rho,z)\over \partial \rho \right|_{\rho=\rho_k}=0,
\label{glstcond}
\end{equation}
for all values of $z$.

As input parameters we use the coefficients of the small-$z$ expansion
of $B(z)$, i.e. the estimates of coefficients $r_{2j}$ obtained by
TM+CFT and reported in Table~\ref{sumrj}.

In Table \ref{eqst1} we report the universal amplitude ratios
derived from truncations corresponding to $k=3,4,5$.
We will not report the results for $k=6$ obtained using the
available estimate of $r_{12}$, because
the relatively large uncertainty on $r_{12}$ induces
a very large error in the results of the $k=6$ truncation.
We only mention that results with $k=6$ are  perfectly consistent with 
those obtained from the $k=5$ truncation.
This can be inferred from the fact that the estimate
of $r_{12}$ obtained using $h^{(5)}(\theta)$, $r_{12}\simeq 4215$, 
is very close to the central value of the TM+CFT estimate,
i.e. $r_{12}=4200(740)$.
As we shall see, we will obtain a
much better estimate of $r_{12}$ in Sec.~\ref{impapp}.

The results of Table~\ref{eqst1} are not stable as $k$ increases, 
showing a systematic
drift up to $k=6$, where the large uncertainty does not allow
a meaningful comparison.
We observe that the results for the universal amplitude ratios
$B_0^\infty$, $R_4^+$, $R_\chi$,
$U_2$ and $v_3$ effectively converge towards their precise estimates
reported in Table~\ref{eqstd2}.
It is also reassuring that the difference between the exact value
and the estimate obtained using $h^{(5)}$
is of the order of the variation of the estimates with changing $k$.

\begin{table}[tbp]
\footnotesize
\begin{center}
\begin{tabular}{|cccc|}
\hline
\multicolumn{1}{|c}{}&
\multicolumn{1}{c}{$h^{(3)}$}&
\multicolumn{1}{c}{$h^{(4)}$}&
\multicolumn{1}{c|}{$h^{(5)}$}\\
\hline
$\rho$
& 2.065 & 2.027 & 2.018(3) \\
$\theta_l^2-\theta_0^2$
& 0.183 & 0.177 & 0.173 \\
$r_6$
& $^*$3.67866(5) &  $^*$3.67866(5) &  $^*$3.67866(5) \\
$r_8$
& 24.413 & $^*$26.041(11) & $^*$26.041(11) \\
$r_{10}$
& 249.11 & 277.1(2) & $^*$284.5(2.4) \\
$r_{12}$
& 3513.7(2) & 4066(4) & 4215(49) \\
$B(|z_0|/2)$
&  1.9621 & 1.9666 & 1.9670  \\
$B(|z_0|)$
& 37.160 & 41.655 & 42.808 \\
$B(2|z_0|)$
& 431786 & 538946 & 569182 \\
$B^{\infty}_0$
& 0.4225$\times 10^{-5}$  & 0.5279(8)$\times 10^{-5}$  &
0.557(10)$\times 10^{-5}$  \\
$R_4^+$
& 7.879 & 7.558(2) & 7.47(3) \\
$R_\chi$
& 7.967 & 7.434(4) & 7.23(7) \\
$U_2$
& 48.565 & 44.41(3) & 42.7(6) \\
$v_3$
& 28.009 & 28.756(6) & 29.2(2) \\
\hline
\end{tabular}
\end{center}
\caption{\sl
Universal amplitude ratios obtained by taking different
approximations of the parametric function $h(\theta)$.
The reported ``errors'' are only related to the
uncertainty of the corresponding input parameters.
Numbers marked with an asterisk are inputs, not predictions.
}
\label{eqst1}
\end{table}

\subsection{Improved approximations from constrained polynomials}
\label{impapp}

Although the polynomial approximations we presented in the previous
Section are substantially consistent
with the known results for the amplitude ratios, they do not provide an
accurate approximation of the equation of state.
The convergence appears rather slow,
probably requiring the knowledge of a considerably larger
number of coefficients $r_{2j}$ to substantially improve the results.
Here, we will present an improved approximation scheme that
significantly increases the precision of the results.

The approximation scheme can be improved by constructing
constrained polynomial approximations of $h(\theta)$
that take into account the large-$z$ asymptotic behaviour of $B(z)$:
\begin{equation}
B(z) = B_0^\infty z^\delta \left[ 1 + O(z^{-1/\beta})\right],
\label{asyFz1}
\end{equation}
where the value of $B_0^\infty$ is reported in Eq.~(\ref{f0inf}).
We consider constrained polynomial approximations of the form
\begin{equation}
\bar{h}^{(k)}(\rho,\theta) =
  \theta  + \sum_{i=1}^{k-1} \bar{h}_{2i+1}(\rho) \theta^{2i+1} +
\bar{h}_{2k+1}(\rho) \theta^{2k+1},
\label{hexpn2}
\end{equation}
where  the coefficients
$\bar{h}_{2i+1}(\rho)$ with $i< k$ are determined as before, by matching
the small-$z$ expansion
of $B(z)$ to $O(z^{2k-1})$, while $\bar{h}_{2k+1}(\rho)$ is fixed by
requiring that
\begin{equation}
B_0^\infty= \rho^{1-\delta} \bar{h}^{(k)}(\rho,1) = 0.592357(6) \times 10^{-5}.
\label{const}
\end{equation}
It follows
\begin{equation}
\bar{h}_{2k+1}(\rho) = \rho^{\delta-1} B^\infty_0 - 1 - \sum_{i=1}^{k-1}
\bar{h}_{2i+1}(\rho).
\label{ht}
\end{equation}
In this approximation scheme the free parameter $\rho$ can be still
determined by requiring the global stationarity condition (\ref{glstcond}).
This nontrivial property
is essentially due to the fact that the constraint (\ref{const}) is linear
in the coefficients
$\bar{h}_{2i+1}$. It can be proved by using arguments similar to
those employed in App. C of
Ref.~\cite{CPRV-99} to show the global stationarity condition (\ref{glstcond})
for the approximation scheme
(\ref{hexpn}).
In Table~\ref{trht2d}, for $k=2,3,4,5$,
we report the polynomials $\bar{h}^{(k)}(\theta)$
obtained  by using the global stationarity condition to fix $\rho$,
and the central values of the input parameters
$F^\infty_0$, $r_6$, $r_8$, $r_{10}$.
Note the stability of the coefficients of the polynomials with $k$ and
that the size of the coefficients decreases with the order of the
polynomial.  The results for some universal quantities
are reported in Table~\ref{eqst2}.
They are in  much better agreement with the exact results
than those obtained without constraint.

\begin{table}[t]
\vspace{0cm}
\footnotesize
\hspace*{0cm}    
\tabcolsep 4pt        
\doublerulesep 1.5pt  
\begin{center}
\begin{tabular}{|c|c|c|l|}
\hline
& & & \\[-3.5mm]
$k$ & $\rho_k$ & $\theta_0^2$ &  $\qquad\qquad\quad
 \bar{h}^{(k)}(\theta)/[\theta (1 - \theta^2/\theta_0^2)]$ \\
\hline
2  &  2.01116 &  1.15278 & $ 1 - 0.208408 \theta^2 $\\
3  &  2.00770 &  1.15940 & $ 1 - 0.215675 \theta^2 - 0.039403 \theta^4 $\\
4  &  2.00770 &  1.16441 & $ 1 - 0.219388 \theta^2 - 0.041791 \theta^4  -
                 0.013488 \theta^6 $\\
5  &  2.00881 &  1.16951 & $ 1 - 0.222389 \theta^2 - 0.043547 \theta^4  -
                 0.014809 \theta^6  - 0.007168 \theta^8 $\\
\hline
\end{tabular}
\end{center}
\caption{\sl
Polynomial approximations of $h(\theta)$ using the global stationarity
condition for various values of the truncation parameter $k$,
cf. Eq.~(\ref{hexpn2}).
The reported expressions correspond to the central values
of the input parameters.
}
\label{trht2d}
\end{table}

\begin{table}[t]
\footnotesize
\begin{center}
\begin{tabular}{|ccccc|}
\hline
& & & \\[-3.5mm]
\multicolumn{1}{|c}{}&
\multicolumn{1}{c}{$\bar{h}^{(2)}$}&
\multicolumn{1}{c}{$\bar{h}^{(3)}$}&
\multicolumn{1}{c}{$\bar{h}^{(4)}$}&
\multicolumn{1}{c|}{$\bar{h}^{(5)}$}\\
\hline
$\rho$
& 2.011 & 2.008 & 2.008 & 2.009(1)  \\
$\theta_l^2-\theta_0^2$
& 0.181 & 0.174 & 0.169 & 0.164(2)  \\
$r_6$
& 3.929 & $^*$3.67866(5) &  $^*$3.67866(5) &  $^*$3.67866(5) \\
$r_8$
& 27.585 & 26.932 & $^*$26.041(11) & $^*$26.041(11) \\
$r_{10}$
& 297.25 & 292.89 &  292.89 & $^*$284.5(2.4) \\
$r_{12}$
& 4425.2 & 4385.6 &  4385.6 & 4443(16) \\
$r_{14}$
& 84387 & 84029 &  84029 & 84305(79) \\
$B(|z_0|/2)$
&  1.9798 & 1.9690 &  1.9675 & 1.9672  \\
$B(|z_0|)$
& 44.930 & 44.335 &  44.146(2) & 44.05(3) \\
$B(3|z_0|/2)$
& 8442.2 & 8432.7  &  8429.4 & 8427.7(5) \\
$B(2|z_0|)$
& 604619(6) & 604548(6) &  604524(6) & 604511(7) \\
$B(3|z_0|)$
& 2.63497(3)$\times 10^8$ & 2.63496(3)$\times 10^8$ & 2.63495(3)$\times
10^8$ & 2.63495(3)$\times 10^8$ \\
$B^{\infty}_0$
& $^*$0.592357(6)$\times 10^{-5}$  & $^*$0.592357(6)$\times 10^{-5}$  &
$^*$0.592357(6)$\times 10^{-5}$  &
$^*$0.592357(6)$\times 10^{-5}$  \\
$B^{\infty}_1$
& 0.021893  & 0.021375 & 0.021198(3) & 0.02110(3)  \\
$B^{\infty}_2$
& 9.3987  & 9.2611 & 9.2611 & 9.286(7) \\
$R_4^+$
& 7.458 & 7.396 & 7.371 & 7.355(5) \\
$R_\chi$
& 7.602 & 7.172 & 7.002(2) & 6.90(3) \\
$U_2$
& 45.918 & 42.358 & 40.76(2)  & 39.6(3)  \\
$v_3$
& 28.328 & 29.201 & 29.837(9) & 30.5(2) \\
\hline
\end{tabular}
\end{center}
\caption{\sl
Universal amplitude ratios obtained from the constrained polynomial
approximations
(\ref{hexpn2}) of the parametric function $h(\theta)$.
The reported ``errors'' are only related to the
uncertainty of the corresponding input parameters
(they are reported only if they are larger than the last figure shown).
Numbers marked
with an asterisk are inputs, not predictions.
}
\label{eqst2}
\end{table}

In  Fig.~\ref{figFzI2d} we show the scaling function
$B(z)$ obtained from $\bar{h}^{(k)}(\rho,\theta)$ for $k=2,3,4,5$.
The convergence is good, indeed their differences are not visible in
Fig.~\ref{figFzI2d}.
This allows us to determine $B(z)$
for all real $z>0$ with a relative precision of at least a few per mille
(the least precision is found around $z\simeq |z_0|\simeq 2.71$).
This fact is not trivial since the small-$z$ expansion has a finite
convergence radius which is expected to be
$|z_0|=(R_4^+)^{1/2}\simeq 2.71$. Therefore,  the determination of $B(z)$
on the whole positive real axis
from its small-$z$ expansion requires an analytic continuation, which is
effectively performed by the approximate parametric representations
we have considered.
We recall that the large-$z$ limit corresponds to the critical isotherm $t=0$,
so that positive real values of $z$ describe the HT phase up to
$t=0$.
Note also the good agreement of the results for $B_2^\infty$
(see Table~\ref{eqst2}), i.e. the next-next-to-leading coefficient
of the large-$z$ expansion of $B(z)$, with the precise estimate
given in Eq.~(\ref{F2infty-Ising2d}).

\begin{figure}[tb]
\hspace{-1cm}
\vspace{0cm}
\centerline{\psfig{width=12truecm,angle=-90,file=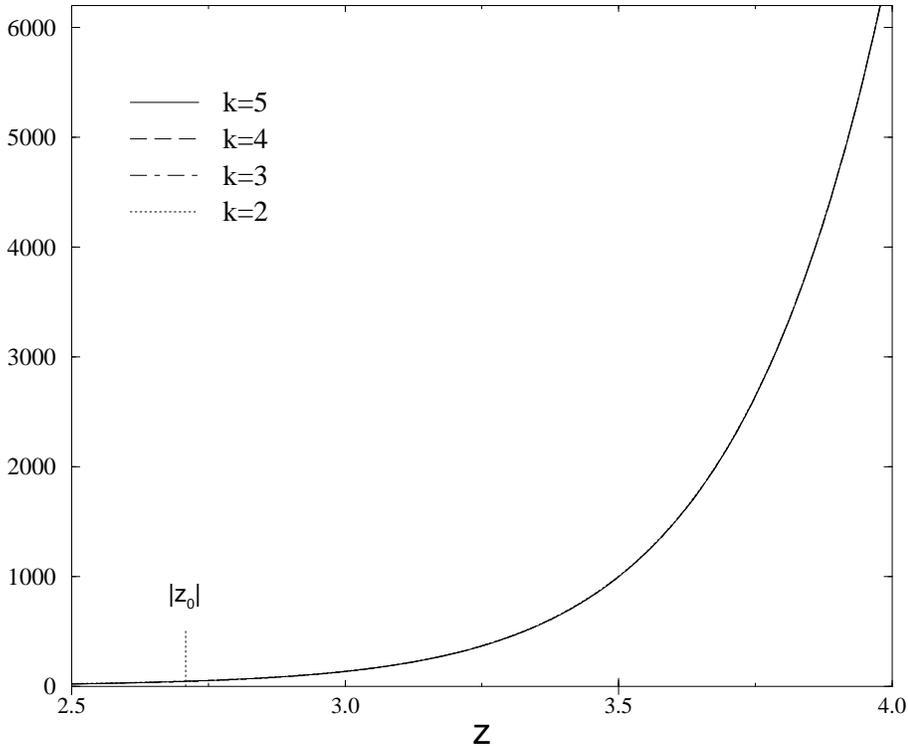}}
\vspace{0cm}
\caption{\sl
The scaling function $B(z)$ versus $z$. We report the curves obtained
from the constrained polynomial approximations (\ref{hexpn2}) with $k=2,3,4,5$.
Their differences are not visible.
}
\label{figFzI2d}
\end{figure}

The convergence of the polynomial representations at the coexistence
curve is slower.  This can be seen by looking at the estimates reported in
Table \ref{eqst2} for universal amplitude ratios involving quantities related
to the coexistence curve, such as $R_4^+$, $R_\chi$, and $U_2$, and
comparing them with the corresponding known results reported in Tables
\ref{eqstd2}. They appear to (monotonically) converge toward
the correct results.
The rate of convergence worsens when quantities with more and more
derivatives with respect to $h$ are involved in the amplitude ratio,
as it can be already seen
by comparing the results for $R_4^+$ and $U_2$.

Fig.~\ref{figfxI2d} shows the scaling function $f(x)$ as obtained by
the truncations $k=2,3,4,5$.
The accuracy of the determination of $f(x)$ can be inferred from
the convergence of the curves with increasing $k$ and the comparison
with the known behavior for $x\to -1$ and $x\to +\infty$.
Indeed, for $x \to -1$ we have 
\begin{eqnarray}
&&f(x) = b_{f} ( 1 + x ) + O\left[ (1+x)^2\right], \\
&&b_f = {\beta U_2\over R_\chi} = 0.69511778...,\nonumber
\end{eqnarray}
and for $x\to +\infty$
\begin{eqnarray}
&&f(x) = f_0^\infty x^\gamma  + O\left( x^{\gamma-2\beta} \right), \\
&&f_0^\infty = R_\chi^{-1} =0.14752994....\nonumber
\end{eqnarray}
This shows that $f(x)$ is determined with a precision of a few per cent
in the whole region.
This happens also in the large-$x$ region, which
corresponds to the HT phase, and therefore  to $z\ll 1$ in $F(z)$,
essentially because $f(x)$ is normalized at the coexistence curve, i.e. $x=-1$,
where our approximation is worse.
For $x>0$, the error on $f(x)$
increases from 0 to 2\%, the relative error on $R_\chi$.

Using the results of Table~\ref{eqst2} we also obtain
\begin{eqnarray}
&&r_{12} = 4.44(6) \times 10^3,        \label{r12est2} \\
&&r_{14} = 8.43(3) \times 10^5,\\
&& B_1^\infty = 0.0211(2).
\end{eqnarray}
Note that the above-reported estimate of $r_{12}$ is perfectly consistent
with the result obtained by TM+CFT, i.e. $r_{12}=4.20(74)\times 10^3$,
but much more precise.

\begin{figure}[tb]
\hspace{-1cm}
\vspace{0cm}
\centerline{\psfig{width=12truecm,angle=-90,file=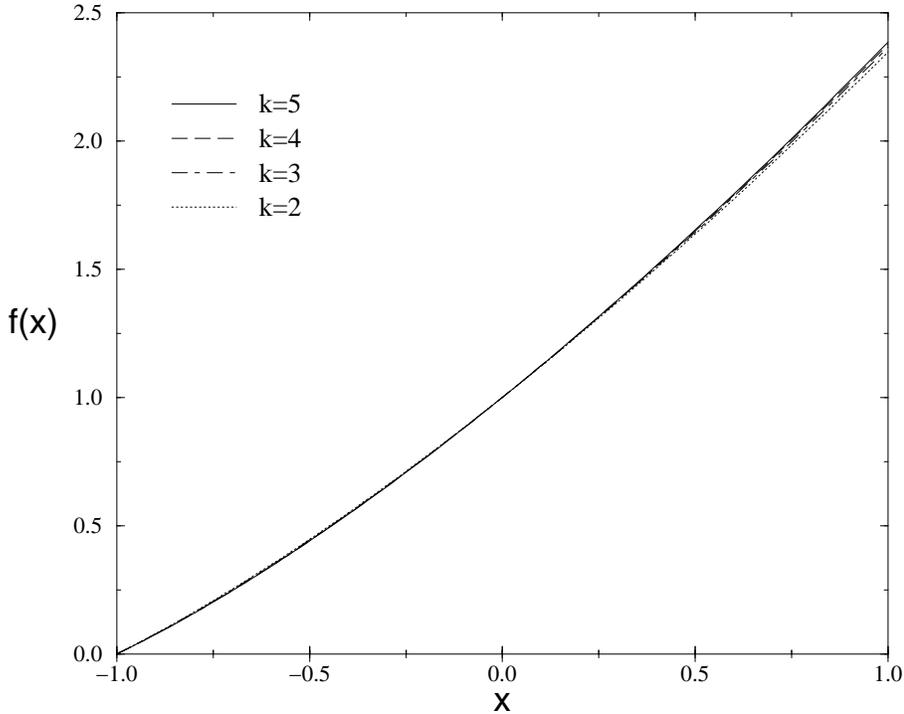}}
\vspace{0cm}
\caption{\sl
The scaling function $f(x)$ versus $x$. We report the curves obtained
from the constrained polynomial
approximations (\ref{hexpn2}) with $k=2,3,4,5$.
}
\label{figfxI2d}
\end{figure}

\section{Conclusions}

Let us briefly summarize the main results of this paper,
which presents a determination of the equation of state of the 
two-dimensional Ising model
in the whole $(t,h)$ plane.
The starting point of the analysis was the determination of the
$2n$-point couplings constants in the HT phase which are
related to the expansion of the free energy in powers of the magnetization.
These quantities  were  then used as input parameters for
a systematic approximation scheme, which allowed us to obtain an
accurate determination of  the equation of
state in the critical regime in the whole $(t,h)$ plane.

The determination of the coefficients $r_{2j}$ of the small-magnetization
expansion of the Helmholtz free energy is in itself an
interesting and nontrivial problem. We addressed this problem by using
TM techniques. In order to overcome the typical problem of all TM
calculations, i.e. the fact that they are constrained to rather small values of
the lattice size in the transverse direction (in our case we could reach
$L_1=24$ as maximum value) we used a two-step strategy:
\begin{description}
\item{1]}
In the first step we studied the
model at very small values of $\beta$ so as to have small values for $\xi$
(whose maximum value was $\xi=3.350288..$,
which was reached for the largest value of
$\beta$ that we studied: $\beta=0.37$) and thus large values of the ratio
$L_1/\xi$. We then extrapolated to the thermodynamic limit the
TM results by using a new effective iterative algorithm.
This algorithm is rather interesting in itself and could be of general
utility for people working with TM methods.

\item{2]}
 Second, we used a combination of standard RG
results and a set of high-precision
 numerical informations on the zero-field magnetic susceptibility to construct
 the scaling functions for  the derivatives of the free energy
involved in the construction of the $r_{2j}$ coefficients. With these scaling
functions we could obtain the critical-limit values of the
 $r_{2j}$ coefficients with rather small errors even if the input data for this
 continuum-limit extrapolation correspond to rather small values of $\beta$.

\end{description}

The expansion of the free energy in powers of the magnetization
in the HT phase was then used as a starting point to construct approximate
representations of the equation of state that are valid in the
critical  regime of the whole $(t,h)$ plane.
We considered a systematic approximation
scheme based on polynomial parametric representations,
devised to match the known terms of the small-magnetization expansion
and the large-magnetization behaviour of the Helmholtz free energy.
A global stationarity condition was used in order to optimize
the polynomial approximation.
This approximation scheme can be improved systematically
by considering higher and higher order polynomials. It is
only limited by the number of known terms in the
small-magnetization expansion of the free energy.
The knowledge of this expansion up to 10th order allowed us to
obtain an accurate determination of the critical equation of state.
We indeed obtained the scaling function $B(z)$, cf. Eq.~(\ref{eqa}),
for all real $z>0$ with a relative precision of at least a few per
mille, and the scaling function $f(x)$, cf. Eq.~(\ref{eqstfx}),
with a precision of a few per cent in the whole physical region $x \geq -1$.
The approximation scheme is systematic, thus this precision can be
improved by a more accurate knowledge
of the small-magnetization expansion of the free energy in the HT phase.

The method that we used to reconstruct the equation of state from the
small-magnetization expansion of the Helmholtz free energy
is general and can be applied to other statistical models.
We mention that similar methods
have been successfully applied to the three-dimensional
Ising ~\cite{GZ-97,CPRV-99} and $XY$ universality~\cite{CPRV-eqst,CHPRV-01} classes,
leading to accurate determinations of
the critical equation of state and of the universal ratios of amplitudes
that can be extracted from it.

\section*{Acknowledgements}
We thank Tony Guttmann for useful discussions and for communicating
 to us the results
collected in~\cite{Guttmann_private} before publication.


\newpage

\appendix

\section{Notations for the amplitudes} \label{AppA}

Universal amplitude ratios characterize the critical behaviour
of thermodynamic quantities that do not depend on
the normalizations of the external (e.g.\ magnetic) field, order
parameter (e.g.\ magnetization) and temperature.  Amplitude ratios of
zero-momentum quantities can be derived from the critical equation of
state. Beside the amplitudes $C^+_{n}$ of the
$n$-point functions $\chi_{n}(t)$
defined in Eq. (\ref{defCpiun}) and the corresponding
ones $C^-_{n}$ in the low-temperature phase, we consider amplitudes derived
from the singular behaviour of the specific heat
\begin{equation}
C_H = A^\pm \ln (1/|t|) ,
\label{sphamp}
\end{equation}
the spontaneous magnetization on the coexistence curve
\begin{equation}
M = B (-t)^{\beta}.
\label{magamp}
\end{equation}
We complete our list of amplitudes by
considering the second-moment correlation length
\begin{equation}
\xi_{\rm 2nd} = f^{\pm} |t|^{-\nu},
\label{xiamp}
\end{equation}
and the true (on-shell) correlation length,
describing the large-distance behaviour of the two-point function,
\begin{equation}
\xi_{\rm gap} = f_{\rm gap}^{\pm} |t|^{-\nu}.
\label{xiosamp}
\end{equation}
One can also define amplitudes along the critical isotherm, e.g.,\
\begin{eqnarray}
\chi &=& C^c |h|^{-{\gamma/\beta\delta}}, \label{chicris}\\
\xi &=& f^c |h|^{-{\nu/\beta\delta}}, \label{xicris}\\
\xi_{\rm gap} &=& f_{\rm gap}^c |h|^{-{\nu/\beta\delta}}. \label{xigapcris}
\end{eqnarray}

\section{Finite-size scaling of the free energy}
\label{appb}
Before starting the discussion on the FSS in the Ising model,
let us stress that
the analysis reported in this appendix is  only
 a straightforward application of results which are already well-known in the
 literature. They can be found for instance in~\cite{km} and are
  based on the results of~\cite{ff} and on the
exact solution of the Ising model on a finite lattice obtained by
Kaufman~\cite{kau}.  We decided to write it down in this appendix all the same
 so as to  make this paper as
self-contained as possible.
We shall use the same notations as ~\cite{km} to simplify the comparison.
This means in particular that we shall use the letter $R$ to denote
the finite size of the lattice in the transverse direction,
which is denoted in the rest of the paper with $L_1$.

The aim of this appendix is to obtain the explicit form of the
FSS of the free energy of the
Ising model on a rectangular lattice of size $R\times\infty$, which is exactly
the geometry we considered in our TM work.
To obtain this result we shall work in the
framework of the S-matrix approach to 2d integrable models and shall in
particular use the so-called Thermodynamic Bethe Ansatz (TBA).
 In the TBA one looks at the theory defined on an
infinitely long  cylinder with circumference $R$. The only free parameter of the
theory is $r\equiv mR$ where $m$ is the lowest mass of the model. The goal is to
extract the behaviour of the free energy $E_0(R)$ and of the lowest mass
$m(R)$ (and possibly, in models more complicated than the Ising one, of higher
massive states) as a function of $R$.
The geometry of the TBA is exactly the one which we have in our TM
setting. The parameter $r$ is the ratio $L/\xi$ in our notations and the only
thing that we must require to compare our findings with the TBA analysis
is that we should be in the ``scaling region,'' i.e.
that $\xi\gg 1$ so that we may neglect the lattice artifacts.

The TBA results depend on the entries of the S-matrix. They are in general
very complicated. Thus it is usually impossible to find the exact FSS functions
for any value of $r$ (one has usually to resort to
  perturbative expansions for  small or
large values of $r$). However, the thermal perturbation of the Ising model is so
simple (the S matrix is simply $S=-1$ as a consequence of the fact that the
model can be mapped to a free massive field theory)
that in this case the explicit expression for any value of
$r$ can be obtained.

\subsection{Free energy at $h=0$}

The TBA prediction for the FSS behaviour of the free energy, in the HT
phase of the 2d Ising model
at $h=0$, is\footnote{Notice that in~\cite{km} the authors study, instead of the
free energy $F(R)$, the quantity  $E_0(R)$ that is related to $F(R)$ by
 $E_0(R)=R[F(R)-F(\infty)]$.}

\eq
F(R)=F(\infty)-\frac{\pi~c(r)}{6R^2}
\label{eq2}
\en
where, following the standard TBA machinery
(see p. 668 of Ref. \cite{km}), $c(r)$ is given by
\eq
c(r)=\frac{6}{\pi^2}\int_0^\infty d\theta \; r \cosh (\theta) 
\ln(1 +e^{-r\cosh \theta})\; .
\label{eq2bis}
\en
The integral can be exactly evaluated and gives:
\eq
c(r)=\frac{6r}{\pi^2}\sum_{k=1}^{\infty}\frac{(-1)^{k-1}}{k} K_1(kr),
\label{eq3}
\en
where $K_1$ is the modified Bessel function of the second kind.

The normalization of Eq. (\ref{eq2}) could seem strange, but it is chosen such
that in the $r\to0$ limit (i.e. at the critical point) the function $c(r)$
exactly becomes the central charge of the Ising model $c=1/2$.
The idea behind this choice is the so called $c$-theorem of Zamolodchikov
which states that the FSS of any
2d integrable model  can be parametrized by Eq.(\ref{eq2}), where $c(r)$ is a
``running central charge,'' which in general
interpolates between two critical points and, in this case, between the central
charge of the Ising model and the value $c(\infty)=0$ which is the appropriate
one for a massive theory.

The limit in which we are interested is $r\gg 1$. In this limit, we can use the
asymptotic expansion for the Bessel functions and Eqs.
(\ref{eq2}), (\ref{eq3}) become:
\eq
F(R)=F(\infty)-\sqrt{\frac{m}{2\pi R^3}}
\sum_{k=1}^{\infty} e^{-mkR}  \frac{(-1)^{k-1}}{k\sqrt{k}} \left(1 + \cdots\right)
\label{eq4}
\en
where the dots stand for the $o(1)$ terms.

\subsection{Derivatives of the free energy at $h=0$}

Since we are interested in the derivatives with respect to $h$ of the free
energy, we must try to extend Eq.~(\ref{eq4}) in the $h\not=0$ plane.
 It is quite reasonable to assume that in  (\ref{eq4}) the
$h$ dependence can only be hidden in the mass term. Moreover, we know from
Refs.~\cite{mw,dms} that this dependence (for small values of $h$) is well
described by an analytic even function of $h$, i.e.
\eq
m(h)=m(1+a~h^2+b~h^4+\ldots ) ,
\en
with $a$ and $b$ unknown constants.

Inserting this expression in Eq.~(\ref{eq4}) and performing the derivatives we
find
\eq
\chi_{2n}(R)-\chi_{2n}(\infty)=R^{2n-3/2}(g_{2n,1}(1/R)e^{-mR}
+g_{2n,2}(1/R)e^{-2mR}
+g_{2n,3}(1/R)e^{-3mR}+...)
\label{eq5}
\en
where the $g_{2n,i}(1/R)$ may be expanded in positive powers of $1/R$
(whose coefficients could be in
principle computed as functions of $a$, $b$, and $m$). This is the expression
quoted in the text as Eq. (\ref{ne1}).

\subsection{A test of the iterative algorithm for the 
infinite-volume extrapolation}
\label{testia}

In order to further check the IA (\ref{eq1}) for the 
infinite-size extrapolation, we apply it to a test function
of the type (\ref{eq5}). We consider the function
\begin{equation}
E(\xi,R) = 1 + {\pi\over 6 R^2}
\left. \partial^4 c(r)\over \partial h^4 \right|_{h=0},
\label{testfunc}
\end{equation}
where $c(r)$ is the function defined in Eq.~(\ref{eq3})
and 
\begin{equation}
r = {R\over \xi} \left( 1 + h^2 + h^4\right).
\end{equation}
Clearly, for $R\to \infty$, $E(\xi,\infty)=1$.
We compute $E(\xi,R)$ for finite values of $R$ in the typical range
of our TM calculations, i.e. $3\ltapprox R/\xi \ltapprox
7$, and apply the IA (\ref{eq1}) to determine
the $E(\xi,\infty)$.  The comparison with the exact value gives
an idea of the effectiveness of the method.
Table~\ref{iatest} shows the results of the IA
(\ref{eq1}) for $\xi=4$, analogously to Table~\ref{iatab} for $F_4$.
We report the results for various iteration levels (up to four);
the level zero corresponds to the original data.
In order to obtain from Table~\ref{iatest} an estimate $E(\xi,\infty)$, we
follow the same strategy used in Sec.~\ref{therlim},
i.e. we consider the largest value of $R$ and four iterations 
of the IA.  The error is estimated
from the residual dependence on $R$ of the results obtained with four 
iterations of the algorithm.
We obtain $E(\xi=4,\infty)=1.000000(1)$,
which is in perfect agreement with the exact number. 
Analogous  results are obtained for other values of $\xi$.
This confirms the effectiveness of the procedure we used to 
perform the extrapolation to  the thermodynamic  limit, and 
the reliability of the uncertainty we considered.

\begin{table}
\footnotesize
\begin{center}
\begin{tabular}{|c|c|c|c|c|c|}
\hline
$R$ &$E^{(0)} $ &$E^{(1)}$ & $E^{(2)} $& $E^{(3)}$ & $E^{(4)}$ \\
\hline
12 & 1.002919604 & $$ & $$ & $$ & $$ \\
13 & 1.0038256116 & $$ & $$ & $$ & $$ \\
14 & 1.0041103404 & 1.0042408301  & $$ & $$ & $$ \\
15 & 1.0040272017 & 1.0040459912  & $$ & $$ & $$ \\
16 & 1.0037413786 & 1.0041444428  & 1.0041113946 & $$ & $$ \\
17 & 1.0033581849 & 1.0048662118  & 1.0040304410 & $$ & $$ \\
18 & 1.0029428048 & 1.0083034836  & 1.0039525998 & 1.0020057776 & $$ \\
19 & 1.0025338931 & 0.9766831739  & 1.0052032241 & 1.0040258799 & $$ \\
20 & 1.0021528510 & 0.9969431584  & 0.9890313495 & 1.0040423723 & 1.0040425080 \\
21 & 1.0018100907 & 0.9987411366  & 0.9989162382 & 0.9951663087 & 1.0040259105 \\
22 & 1.0015092173 & 0.9993480457  & 0.9996572961 & 0.9997173548 & 0.9981748024 \\
23 & 1.0012497886 & 0.9996258634  & 0.9998603960 & 0.9999370740 & 0.9999482199 \\
24 & 1.0010291081 & 0.9997722737  & 0.9999353998 & 0.9999793168 & 0.9999893714 \\
25 & 1.0008433668 & 0.9998559476  & 0.9999675467 & 0.9999916598 & 0.9999967552 \\
26 & 1.0006883533 & 0.9999063496  & 0.9999827009 & 0.9999962157 & 0.9999988812 \\
27 & 1.0005598763 & 0.9999378507  & 0.9999903522 & 0.9999981547 & 0.9999995914 \\
28 & 1.0004540024 & 0.9999580850  & 0.9999944245 & 0.9999990581 & 0.9999998462 \\
\hline
\end{tabular}
\end{center}
\caption{\sl Results of the IA (\ref{eq1}) applied to the test function
$E(\xi=4,R)$, cf. Eq.~(\ref{testfunc}).
}
\label{iatest}
\end{table}

\section{Determination of $\lowercase{e}_{\lowercase{h}}$}
\label{appc}

In this appendix we present the determination of $e_h$. We will obtain
it by analyzing the scaling corrections to the free energy in the presence
of a magnetic field $h$ on the critical isotherm $t=0$.
The scaling corrections for $h\to 0$ have been determined in~\cite{ch99}.
Using the notation of~\cite{ch99},
we have, cf. Eq. (84) of ref.~\cite{ch99}:
\eq
F(0,h)=~ f_b~+  A^l_f~|h|^{\frac{16}{15}}(1+
A^l_{f,b}|h|^{\frac{14}{15}}+
A^l_{f,1}|h|^{\frac{16}{15}}+
A^l_{f,2}|h|^{\frac{22}{15}}+
A^l_{f,3}|h|^{\frac{30}{15}}+ \cdots ).
\label{ac2}
\en
Each of these amplitudes has a precise physical meaning. Let us look at them
in detail:
\begin{description}

\item{\bf $f_b$}  denotes the bulk contribution to the free energy and
can be obtained from the
exact solution of the Ising model on the lattice at the critical point.
Explicitly
$f_b = \case{2}{\pi} G + \case{1}{2}\log 2$, where $G$ is Catalan's constant.

\item{\bf $A^l_{f}$}
is the amplitude of the singular part of the free energy. It can be
evaluated exactly in the framework of the S-matrix approach to the model.
Its value is $A^l_{f}=0.9927995...$ \cite{zam,ch99}.

\item{\bf $A^l_{f,b}$}
is the first correction (proportional to $h^2$) due to the bulk part of
the free energy. It can be related to the constant term
$D_0 = -0.104133245\ldots$ \cite{KAP-86}  appearing
in the expansion in powers of $t$ of the susceptibility at $h=0$.
Its value is $A^l_{f,b}=D_0/(2 A^l_{f}) = -0.0524442...$

\item{\bf $A^l_{f,1}$}
is due to the $T\bar T$, $T^2$ and $\bar T^2$ irrelevant
operators in the Hamiltonian. This amplitude turns out to be
compatible with zero~\cite{ch99}.

\item{\bf $A^l_{f,2}$}
is  due to the $b_th^2$ term in $u_t$.

\item{\bf $A^l_{f,3}$}
is  due to the  $e_hh^2$ term in $u_h$.

\end{description}

In principle one could use the TM data to estimate
all these constants. In practice this procedure works only for the first
unknown amplitude. All the higher ones are then ``shadowed'' by the first.

 The problem with $e_h$ is that  it appears at a
rather high level. In~\cite{ch99} we were able to fix exactly
the amplitudes only up to $A^l_{f,1}$. The first unknown one was
$A^l_{f,2}$ which was then estimated numerically, obtaining
\eq
0.020~ <~ A^l_{f,2}~ <~ 0.022 \;\;\;.
\en
It was impossible to give any reliable estimate for
$A^l_{f,3}$. The main progress of the present
 paper with respect to that analysis
 is that, thanks to the exact
calculation of $b_t$ performed in~\cite{Nickel,CHPV-00}
we are now in the position to
estimate exactly also $A^l_{f,2}$. A direct calculation gives
\eq
 A^l_{f,2}= \frac{A^l_E \pi E_0}{A^l_f 8 \beta_c},
\label{ac1}
\en
where $A^l_f$ was define above, $E_0$ is given by Eq. (\ref{E0}),
and $A^l_E$ is defined by the singular behaviour of the internal energy,
i.e. $E_{\rm int}(t=0,h) = \case{1}{2} \partial F/\partial\beta =
      E_{\rm bulk} +
      A^l_E h^{8/15} + \ldots$ Numerically,
      $A^l_E=0.58051...$ \cite{FLZZ,ch99}.
Substituting in Eq. (\ref{ac1}), we find
\eq
A_{f,2}=0.0210115...
\en
in perfect agreement with the estimate of~\cite{ch99}.

Substituting in Eq. (\ref{ac2}),
we may now estimate numerically the
amplitude $A_{f,3}^l$, which is related to $e_h$ by:
\eq
  A_{f,3}=\frac{16}{15} e_h .
\en
A standard application of the fitting procedure discussed in~\cite{ch99}, using
as input data those reported in Table 10 of~\cite{ch99} gives
\eq
      e_h= -0.00727(15).
\en

\end{document}